\pdfoutput=1
\documentclass[twocolumn,10pt,twoside]{IEEEtran}
\usepackage{xcolor}
\usepackage{pdfcolmk}
\usepackage{float}
\usepackage{units}
\usepackage{bm}
\usepackage{amsthm}
\usepackage{amsmath}
\usepackage{amssymb}
\usepackage{graphicx}
\usepackage[utf8]{inputenc}
\PassOptionsToPackage{normalem}{ulem}
\usepackage{ulem}
\usepackage[unicode=true,
 bookmarks=true,bookmarksnumbered=true,bookmarksopen=true,bookmarksopenlevel=1,
 breaklinks=false,pdfborder={0 0 0},backref=false,colorlinks=false,hidelinks]
 {hyperref}
\hypersetup{pdftitle={MERLiN: Mixture Effect Recovery in Linear Networks},
 pdfauthor={Sebastian Weichwald, Moritz Grosse-Wentrup, Arthur Gretton},
 pdfpagelayout=OneColumn, pdfnewwindow=true, pdfstartview=XYZ, plainpages=false}

\makeatletter

\floatstyle{ruled}
\newfloat{algorithm}{tbp}{loa}
\providecommand{\algorithmname}{Algorithm}
\floatname{algorithm}{\protect\algorithmname}

\theoremstyle{plain}
\newtheorem{thm}{\protect\theoremname}
\theoremstyle{definition}
\newtheorem{defn}[thm]{\protect\definitionname}
\theoremstyle{definition}
\newtheorem{example}[thm]{\protect\examplename}
\theoremstyle{remark}
\newtheorem{claim}[thm]{\protect\claimname}

\usepackage[caption=false,font=footnotesize]{subfig}

\usepackage{setspace}

\usepackage{pgfplots}
\tikzset{node/.style={circle, draw,font=\sffamily\bfseries, minimum size=2.5em, text centered, line width=1pt}}
\tikzset{nodeh/.style={circle, dashed, draw,font=\sffamily\bfseries, minimum size=2.5em, text centered, line width=1pt}}
\tikzset{d/.style={line width=1pt}}
\usetikzlibrary{snakes}
\tikzset{waves/.style={snake=expanding waves,segment length=2pt,gray}}

\usepackage{amsthm}
\usepackage{amsmath}
\usepackage{centernot}
\def\independenT#1#2{\mathrel{\rlap{$#1#2$}\mkern2mu{#1#2}}}
\newcommand{\indep}{\protect\mathpalette{\protect\independenT}{\perp}}
\newcommand{\dep}{\centernot\indep}

\usepackage{balance}

\makeatother

\providecommand{\claimname}{Claim}
\providecommand{\definitionname}{Definition}
\providecommand{\examplename}{Example}
\providecommand{\theoremname}{Theorem}

\newcommand\copyrighttext{
This is the author's version of an article that is published in \textit{IEEE Journal of Selected
Topics in Signal Processing,} 10(7), 1254--1266, 2016, \href{http://dx.doi.org/10.1109/JSTSP.2016.2601144}{doi: 10.1109/JSTSP.2016.2601144}.
Copyright (c) 2016 IEEE. Personal use is permitted. For any other purposes, permission must be obtained from the IEEE by emailing pubs-permissions@ieee.org.
}

\newcommand\copyrightnotice{%
\begin{tikzpicture}[remember picture,overlay]
\node[anchor=south] at (current page.south) {{\parbox{\dimexpr\textwidth-\fboxsep-\fboxrule\relax}{\centering\scriptsize\copyrighttext}}};
\end{tikzpicture}%
}

\begin{document}

\title{\centerline{MERLiN: Mixture Effect Recovery in Linear Networks}}

\author{Sebastian Weichwald, Moritz Grosse-Wentrup, Arthur Gretton%
\thanks{%
Sebastian Weichwald is with the Empirical Inference Department, Max
Planck Institute for Intelligent Systems, Tübingen, Germany, and has
been with the Centre for Computational Statistics and Machine Learning,
University College London, London, United Kingdom, e-mail: \protect\href{mailto:sweichwald@tue.mpg.de}{sweichwald@tue.mpg.de}.%
}%
\thanks{Moritz Grosse-Wentrup is with the Empirical Inference Department,
Max Planck Institute for Intelligent Systems, Tübingen, Germany, e-mail:
\protect\href{mailto:moritzgw@tuebingen.mpg.de}{moritzgw@tuebingen.mpg.de}.%
}%
\thanks{Arthur Gretton is with the Gatsby Computational Neuroscience Unit,
Sainsbury Wellcome Centre, London, United Kingdom, e-mail: \protect\href{mailto:arthur.gretton@gmail.com}{arthur.gretton@gmail.com}.%
}}

\maketitle
\copyrightnotice

\begin{abstract}
Causal inference concerns the identification of cause-effect relationships between variables, e.\,g. establishing whether a stimulus affects activity in a certain brain region.
The observed variables themselves often do not constitute meaningful \emph{causal} variables, however, and linear combinations need to be considered.
In electroencephalographic studies, for example, one is not interested in establishing cause-effect relationships between electrode signals (the observed variables), but rather between cortical signals (the causal variables) which can be recovered as linear combinations of electrode signals.

We introduce MERLiN (Mixture Effect Recovery in Linear Networks), a family of causal inference algorithms that implement a novel means of constructing causal variables from non-causal variables.
We demonstrate through application to EEG data how the basic MERLiN algorithm can be extended for application to different (neuroimaging) data modalities.
Given an observed linear mixture, the algorithms can recover a causal variable that is a linear effect of another given variable.
That is, MERLiN allows us to recover a cortical signal that is affected by activity in a certain brain region, while not being a direct effect of the stimulus.
The Python/Matlab implementation for all presented algorithms is available on \href{https://github.com/sweichwald/MERLiN}{https://github.com/sweichwald/MERLiN}.
\end{abstract}


\section{Introduction}

Causal inference requires causal variables.
Observed variables do not themselves always constitute the causal relata that admit meaningful causal statements, however, and transformations of the variables might be required to isolate causal signals.
Images,
for example, consist of microscopic variables (pixel colour values)
while the identification of meaningful cause-effect relationships
requires the construction of macroscopic causal variables (e.\,g. whether
the image shows a magic wand) \cite{chalupka2014visual}.
That is, it is implausible that a description of effects of changing the colour value of one single pixel, the microscopic variable, leads to characterisation of a meaningful cause-effect relationship; however, the existence of a magic wand, the macroscopic variable, may lead to meaningful statements of the form ``manipulating the image such that it shows a magic wand affects the chances of little Maggie favouring this image''.
A similar
problem often occurs whenever only a linear mixture of causal variables
can be observed. In electroencephalography (EEG) studies, for example,
measurements at electrodes placed on the scalp are considered to
be instantaneously and linearly superimposed electromagnetic activity
of sources in the brain \cite{Nunez2006}.
Again, statements about the microscopic variables are meaningless, e.\,g. ``manipulating the electrode's signal affects the subject's attentional state''; however, macroscopic variables such as the activity in the parietal cortex, extracted as a linear combination of electrode signals, admit meaningful causal statements such as ``manipulating the activity in the parietal cortex affects the subject's attentional state''.
Standard causal inference
methods require that all underlying causal variables, i.\,e., the sources in the brain, are
first constructed --or rather recovered-- from the observed mixture, i.\,e., the electrode signals.

There exist a plethora of methods to construct macroscopic variables
both from images and linear mixtures. However, prevailing methods
to learn visual features \cite{lowe1999object,dalal2005histograms,bay2008surf}
ignore the causal aspect, and are fundamentally different from the
recent and (to our knowledge) only work that demonstrates how visual \emph{causal} features
can be learned by a sequence of interventional experiments~\cite{chalupka2014visual}.
Likewise, methods to \mbox{(re-)construct} variables from linear mixtures
commonly ignore the causal aspect and often rest on implausible assumptions.
For instance, independent component analysis (ICA), commonly employed
in the analysis of EEG data, rests on the assumption of mutually independent
sources \cite{Hyvarinen2000,Makeig2002,Hyvarinen2004}. One may argue
that muscular or ocular artefacts are independent of the cortical
sources and can be extracted via ICA \cite{Iriarte2003,Delorme2007}.
It seems implausible, though, that cortical sources are mutually independent.
In fact, if they were mutually independent there would be no cause-effect
relationships between them. Thus, methods
ignoring the causal aspect are unsuited to construct
meaningful causal variables.

Mixture Effect Recovery in Linear Networks (MERLiN) aims to construct
a causal variable from a linear mixture without requiring multiple
interventional experiments. The fundamental idea is to directly search
for statistical in- and dependences that imply, under assumptions
discussed below, a certain cause-effect relationship. In essence,
given iid samples of a univariate randomised variable $S$, a univariate
causal effect $C_{1}$ of $S$, and a multivariate variable $F$,
MERLiN searches for a linear combination $\bm{w}$ such that $\bm{w}^{\top}F$
is a causal effect of $C_{1}$, i.\,e., $S\to C_1 \to \bm{w}^{\top}F$. This implements the novel idea to
construct causal variables such that the resulting statistical properties
guarantee meaningful cause-effect relationships.

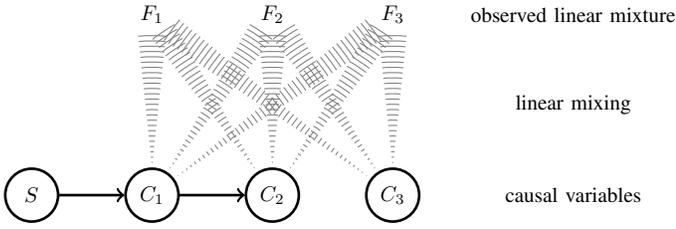
\begin{figure}[tb]
\begin{center}
\-\newline
\begin{tikzpicture}[scale=.8, every node/.style={scale=.8}]
    \node[node] (s) at(0,0) {$S$};

    \node[node] (c1) at(2,0) {$C_1$};
    \node[node] (c2) at(4,0) {$C_2$};
    \node[node] (c3) at(6,0) {$C_3$};

    \node (f1) at(2,3) {$F_1$};
    \node (f2) at(4,3) {$F_2$};
    \node (f3) at(6,3) {$F_3$};

    \node (cmt1) at(9,3) {observed linear mixture};
    \node (cmt2) at(9,1.5) {linear mixing};
    \node (cmt3) at(9,0) {causal variables};

    \draw[waves,segment angle=6] (c1) -- (f1);
    \draw[waves,segment angle=4.5] (c1) -- (f2);
    \draw[waves,segment angle=3] (c1) -- (f3);

    \draw[waves,segment angle=4.5] (c2) -- (f1);
    \draw[waves,segment angle=6] (c2) -- (f2);
    \draw[waves,segment angle=4.5] (c2) -- (f3);

    \draw[waves,segment angle=3] (c3) -- (f1);
    \draw[waves,segment angle=4.5] (c3) -- (f2);
    \draw[waves,segment angle=6] (c3) -- (f3);

    \draw[->,d] (s) -- (c1);
    \draw[->,d] (c1) -- (c2);
\end{tikzpicture}\end{center}\caption{Problem illustration where $S\to C_1 \to C_2 \quad C_3$ is the underlying causal graph and $F_1,F_2,F_3$ are observed variables that are a linear mixture of $C_1,C_2,C_3$.}
\label{fig:example-basic}
\end{figure}

As an illustration, consider the directed acyclic graph (DAG) $S \to C_1 \to C_2\quad C_3$ shown
in Figure~\ref{fig:example-basic}, where the gap indicates that $C_{3}$ is disconnected from all other
variables. In this notation edges denote
cause-effect relationships starting at the cause and pointing towards
the effect. $S$ denotes a randomised variable. We assume that only
a linear mixture $F=\bm{A}[C_{1},C_{2},C_{3}]^{\top}$ ($\bm{A}$
is an unknown mixing matrix) of the causal variables $C_{1},C_{2},C_{3}$ can
be observed, and that $\bm{v}$ such that $C_{1}=\bm{v}^{\top}F$ is known. MERLiN's goal is to
recover from the observed linear mixture $F=[F_1,F_2,F_3]^\top$ a causal variable that is an effect of $C_{1}$, i.\,e., to find $\bm{w}$ such
that $\bm{w}^{\top}F$ is an effect of $C_{1}$, where $\bm{w}^{\top}F=C_{2}$
is a valid solution.

We introduce the basic MERLiN$_{\Sigma^{-1}}$ algorithm that can
recover the sought-after causal variable when the cause-effect relationships
are linear with additive Gaussian noise (cf. Section~\ref{sec:basicmerlin}).
We also demonstrate how the algorithm can be extended for application to different (neuroimaging) data modalities by (a) including data specific preprocessing steps into the optimisation procedure, and (b) incorporating a priori domain knowledge (cf. Section~\ref{sec:extendingmerlin}).
Here, both concepts are demonstrated through application to EEG data when cause-effect relationships within individual frequency bands are of interest.

Thus, we present three related algorithms MERLiN$_\Sigma^{-1}$, MERLiN$_{\Sigma^{-1}}^{bp}$, and MERLiN$_{\Sigma^{-1}}^{bp+}$ that are based on optimisation of precision matrix entries (indicated by the subscript $\Sigma^{-1}$).
The MERLiN$_{\Sigma^{-1}}^{bp}$ and MERLiN$_{\Sigma^{-1}}^{bp+}$ algorithms include preprocessing steps that allow application to timeseries data, identifying a linear combination of timeseries signals such that the resulting log-bandpower (indicated by the superscript $bp$) reveals the sough-after cause-effect relationship.
A further extension (indicated by the superscript $bp+$) takes domain knowledge about time-lags into account.

For stimulus-based neuroimaging studies \cite{weichwald2014}, the MERLiN$_{\Sigma^{-1}}^{bp}$
and MERLiN$_{\Sigma^{-1}}^{bp+}$ algorithms can establish a cause-effect
relationship between brain state features that are observed only as
part of a linear mixture. As such, MERLiN is able to provide insights
into brain networks beyond those readily obtained from encoding and
decoding models trained on pre-defined variables \cite{Weichwald2015}.
Furthermore, it employs the framework of Causal Bayesian Networks
(CBNs) that has recently been fruitful in the neuroimaging community
\cite{Ramsey2010,Grosse2011,ramsey2011multi,Mumford2014,Weichwald2015,grosse2015identification}
--- the important advantage over methods based on information flow
being that it yields testable predictions on the impact of interventions
\cite{eichler2010granger,lizier2010differentiating}.

MERLiN shows good performance both on synthetic and EEG data recorded
during neurofeedback experiments. The Python/Matlab implementation
for all presented algorithms is available on \href{https://github.com/sweichwald/MERLiN}{https://github.com/sweichwald/MERLiN}.

\section{Preliminaries}

\subsection{Causal Bayesian Networks}\label{sub:CBNs}

In general causal inference requires three steps.
\begin{enumerate}
    \item construction of (causal) variables
    \item inference of cause-effect relationships among the variables defined in 1)
    \item estimation of the functional form and strength of the causal links established in 2)
\end{enumerate}
In this manuscript we focus on and merge the first two steps.
More specifically, a causal variable is implicitly constructed by optimising for properties that at the same time establish a specific cause-effect relationship for this variable.
Here we briefly introduce the main aspects of Causal Bayesian Networks
(CBNs) that define causation in terms of effects of interventions and allow inference of cause-effect relationships (step 2) from conditional independences in the observed distribution.
For an exhaustive treatment see \cite{Spirtes2000,Pearl2009}.

\begin{defn}[Structural Equation Model]
We define a \emph{structural equation model (SEM) $\mathcal{{S}}$}
as a set of structural equations $X_{i}=f_{i}(\mathbf{{PA}}_{i},N_{i}),\ i\in\mathbb{{N}}_{1:s} \triangleq \{n\in\mathbb{N}: 1\leq n \leq s\}$
where the so-called noise variables are independently distributed
according to $\mathbb{{P}}_{N_{1},...,N_{s}}=\mathbb{{P}}_{N_{1}}\cdots\mathbb{{P}}_{N_{s}}$.
For $i\in\mathbb{{N}}_{1:s}$ the set $\mathbf{{PA}}_{i}\subseteq\{X_{1},...,X_{s}\}\setminus X_{i}$
contains the so-called parents of $X_{i}$ and $f_{i}$ describes
how $X_{i}$ relates to the random variables in $\mathbf{{PA}}_{i}$
and $N_{i}$. This induces the unique joint distribution denoted by $\mathbb{{P}}_{\mathcal{{S}}}\triangleq\mathbb{{P}}_{X_{1},...,X_{s}}$.%
\footnote{Note that the distribution $\mathbb{{P}}_{\mathcal{{S}}}$ has a density if $\mathbb{{P}}_{N_{1},...,N_{s}}$ has a density and the functions $f_i,\ i\in\mathbb{{N}}_{1:s}$ are differentiable.}

Replacing at least one of the functions $f_{i},\ i\in\mathbb{{N}}_{1:s}$
by a constant $\spadesuit$ yields a new SEM. We say $X_{i}$ has
been intervened on, which is denoted by $\operatorname{do}(X_{i}=\spadesuit)$,
leads to the SEM $\mathcal{{S}}|\operatorname{do}(X_{i}=\spadesuit)$,
and induces the \emph{interventional distribution} $\mathbb{{P}}_{\mathcal{{S}}|\operatorname{do}(X_{i}=\spadesuit)}\triangleq\mathbb{{P}}_{X_{1},...,X_{s}|\operatorname{do}(X_{i}=\spadesuit)}$.
\end{defn}

\begin{defn}[Cause and Effect]
\label{def:causeeffect}$X_{i}$ is a \emph{cause} of $X_{j}$ ($i,j\in\mathbb{{N}}_{1:s},\ i\neq j$)
wrt. a SEM $\mathcal{{S}}$ iff there exists $\heartsuit\in\mathbb{{R}}$
such that $\mathbb{{P}}_{X_{j}|\operatorname{do}(X_{i}=\heartsuit)}\neq\mathbb{{P}}_{X_{j}}$.%
\footnote{$\mathbb{{P}}_{X_{j}|\operatorname{do}(X_{i}=\heartsuit)}$ and $\mathbb{{P}}_{X_{j}}$
denote the marginal distributions of $X_{j}$ corresponding to $\mathbb{{P}}_{\mathcal{{S}}|\operatorname{do}(X_{i}=\heartsuit)}$
and $\mathbb{{P}}_{\mathcal{{S}}}$ respectively.%
} $X_{j}$ is an \emph{effect} of $X_{i}$ iff $X_{i}$ is a cause
of $X_{j}$. Often the considered SEM $\mathcal{{S}}$ is omitted
if it is clear from the context.
\end{defn}
For each SEM $\mathcal{{S}}$ there is a corresponding graph $\mathcal{{G_{S}}}(V,E)$
with $V\triangleq\{X_{1},...,X_{s}\}$ and $E\triangleq\{(X_{i},X_{j}):\ X_{i}\in\mathbf{{PA}}_{j},\ X_{j}\in V\}$
that has the random variables as nodes and directed edges pointing
from parents to children. We employ the common assumption that this
graph is acyclic, i.\,e., $\mathcal{{G_{S}}}$ will always be a directed
acyclic graph (DAG).

It is insightful to consider the following implication of Definition
\ref{def:causeeffect}: If in $\mathcal{{G_{S}}}$ there is no directed
path from $X_{i}$ to $X_{j}$, $X_{i}$ is not a cause of $X_{j}$
(wrt. $\mathcal{{S}}$). The following example shows that without
further assumptions the converse is not true in general, i.\,e., existence of a path does not generally imply a cause-effect relationship.
This nuisance will be accounted for by the faithfulness assumption (cf. Definition~\ref{def:faithful} below).
We provide
supportive graphical depictions in Figure \ref{fig:4ex}.
\begin{example}
\label{exa:faithful}Consider a SEM $S$ with structural equations
and graph $\mathcal{{G_{S}}}$ shown in Figure \ref{fig:4ex1} and
noise variables $(N_{1},N_{2},N_{3})\sim\mathcal{{N}}(0,1)^{3}$.
In $\mathcal{{G_{S}}}$ there is a directed path (in fact even a directed
edge) from $X_{1}$ to $X_{3}$ while $\mathbb{{P}}_{X_{3}|\operatorname{do}(X_{1}=\heartsuit)}=\mathbb{{P}}_{X_{3}}=\mathbb{{P}}_{N_{2}+N_{3}}=\mathcal{{N}}(0,2)$
for all $\heartsuit\in\mathbb{{R}}$, i.\,e., intervening on $X_2$ does not alter the distribution of $X_1$.
That is, $X_{1}$ is not a cause of $X_{3}$ wrt. $\mathcal{{S}}$ despite the existence of the edge $X_1 \to X_3$ (cf. Figure \ref{fig:4ex2}).

Observe that $\mathbb{{P}}_{X_{3}|\operatorname{do}(X_{2}=\heartsuit)}=\mathcal{{N}}(\heartsuit,2)\neq\mathcal{{N}}(0,2)=\mathbb{{P}}_{X_{3}}$
for $\heartsuit\neq0$, i.\,e., $X_{2}$ is, as one may intuitively expect,
a cause of $X_{3}$ wrt. $\mathcal{{S}}$ (cf. Figure \ref{fig:4ex3}).
Likewise, $X_{3}$ indeed turns out not to be a cause of $X_{1}$
or $X_{2}$ as can be seen from Figure \ref{fig:4ex4}.

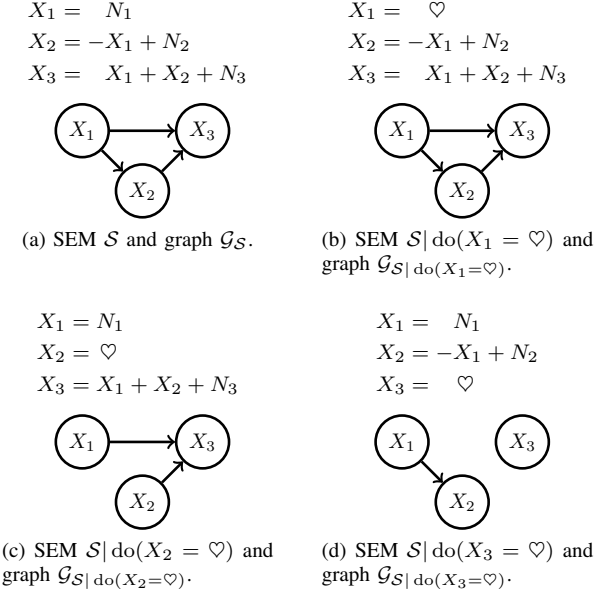
\begin{figure}[tb]
\begin{minipage}[t]{0.48\columnwidth}%
\begin{center}{\footnotesize

\begin{align*}
X_{1} & =\quad\! N_{1}\\
X_{2} & =-X_{1}+N_{2}\\
X_{3} & =\quad\! X_{1}+X_{2}+N_{3}
\end{align*}
}\vspace{-2em}

\subfloat[SEM $\mathcal{{S}}$ and graph $\mathcal{{G_{S}}}$.]{\hspace{-1em}\label{fig:4ex1}

\begin{tikzpicture}[scale=.8, every node/.style={scale=.8}]
	\node (a) at(-1.5,0) {};
	\node (a) at(3.5,0) {};
	\node[node] (X1) at(0,0) {$X_1$};
	\node[node] (X2) at(1,-1) {$X_2$};
	\node[node] (X3) at(2,0) {$X_3$};
    \draw[->,d] (X1) -- (X2);
	\draw[->,d] (X1) -- (X3);
	\draw[->,d] (X2) -- (X3);
\end{tikzpicture}\hspace{-1em}}

\end{center}%
\end{minipage}%
\begin{minipage}[t]{0.48\columnwidth}%
\begin{center}{\footnotesize

\begin{align*}
X_{1} & =\quad\heartsuit\\
X_{2} & =-X_{1}+N_{2}\\
X_{3} & =\quad\! X_{1}+X_{2}+N_{3}
\end{align*}
}\vspace{-2em}

\subfloat[SEM $\mathcal{{S}}|\operatorname{do}(X_{1}=\heartsuit)$ and graph
$\mathcal{{G}}_{\mathcal{{S}}|\operatorname{do}(X_{1}=\heartsuit)}$.]{\hspace{-1em}\label{fig:4ex2}

\begin{tikzpicture}[scale=.8, every node/.style={scale=.8}]
	\node (a) at(-1.5,0) {};
	\node (a) at(3.5,0) {};
	\node[node] (X1) at(0,0) {$X_1$};
	\node[node] (X2) at(1,-1) {$X_2$};
	\node[node] (X3) at(2,0) {$X_3$};
    \draw[->,d] (X1) -- (X2);
	\draw[->,d] (X1) -- (X3);
	\draw[->,d] (X2) -- (X3);
\end{tikzpicture}\hspace{-1em}}

\end{center}%
\end{minipage}

\smallskip{}

\begin{minipage}[t]{0.48\columnwidth}%
\begin{center}{\footnotesize

\begin{align*}
X_{1} & =N_{1}\\
X_{2} & =\,\heartsuit\\
X_{3} & =X_{1}+X_{2}+N_{3}
\end{align*}
}\vspace{-2em}

\subfloat[SEM $\mathcal{{S}}|\operatorname{do}(X_{2}=\heartsuit)$ and graph
$\mathcal{{G}}_{\mathcal{{S}}|\operatorname{do}(X_{2}=\heartsuit)}$.]{\hspace{-1em}\label{fig:4ex3}

\begin{tikzpicture}[scale=.8, every node/.style={scale=.8}]
	\node (a) at(-1.5,0) {};
	\node (a) at(3.5,0) {};
	\node[node] (X1) at(0,0) {$X_1$};
	\node[node] (X2) at(1,-1) {$X_2$};
	\node[node] (X3) at(2,0) {$X_3$};
	\draw[->,d] (X1) -- (X3);
	\draw[->,d] (X2) -- (X3);
\end{tikzpicture}\hspace{-1em}}

\end{center}%
\end{minipage}%
\begin{minipage}[t]{0.48\columnwidth}%
\begin{center}{\footnotesize

\begin{align*}
X_{1} & =\quad\! N_{1}\\
X_{2} & =-X_{1}+N_{2}\\
X_{3} & =\quad\heartsuit
\end{align*}
}\vspace{-2em}

\subfloat[SEM $\mathcal{{S}}|\operatorname{do}(X_{3}=\heartsuit)$ and graph
$\mathcal{{G}}_{\mathcal{{S}}|\operatorname{do}(X_{3}=\heartsuit)}$.]{\hspace{-1em}\label{fig:4ex4}

\begin{tikzpicture}[scale=.8, every node/.style={scale=.8}]
	\node (a) at(-1.5,0) {};
	\node (a) at(3.5,0) {};
	\node[node] (X1) at(0,0) {$X_1$};
	\node[node] (X2) at(1,-1) {$X_2$};
	\node[node] (X3) at(2,0) {$X_3$};
    \draw[->,d] (X1) -- (X2);
\end{tikzpicture}\hspace{-1em}}

\end{center}%
\end{minipage}
\caption{SEMs and graphs accompanying example \ref{exa:faithful}.}
\label{fig:4ex}
\end{figure}

\end{example}
So far a DAG $\mathcal{{G_{S}}}$ simply depicts all parent-child
relationships defined by the SEM $\mathcal{{S}}$. Missing directed
paths indicate missing cause-effect relationships. In order to specify
the link between statistical independence (denoted by $\indep$) wrt.
the joint distribution $\mathbb{{P}}_{\mathcal{{S}}}$ and properties
of the DAG $\mathcal{G_{S}}$ (representing a SEM $\mathcal{S}$)
we need the following definitions.
\begin{defn}[d-separation]
For a fixed graph $\mathcal{{G}}$ disjoint sets of nodes $A$ and
$B$ are \emph{d-separated} by a third disjoint set $C$ (denoted
by $A\perp_{\text{{d-sep}}}B|C$) iff all pairs of nodes $a\in A$
and $b\in B$ are d-separated by $C$. A pair of nodes $a\neq b$
is d-separated by $C$ iff every path between $a$ and $b$ is blocked
by $C$. A path between nodes $a$ and $b$ is blocked by $C$ iff
there is an intermediate node $z$ on the path such that (i) $z\in C$
and $z$ is tail-to-tail ($\gets z\to$) or head-to-tail ($\to z\to$),
or (ii) $z$ is head-to-head ($\to z\gets$) and neither $z$ nor
any of its descendants is in $C$.
\end{defn}

\begin{defn}[Markov property]
A distribution $\mathbb{{P}}_{X_{1},...,X_{s}}$ satisfies the \emph{global
Markov property} wrt a graph $\mathcal{{G}}$ if 
\[
A\perp_{\text{{d-sep}}}B|C\quad\implies\quad A\indep B|C.
\]
It satisfies the \emph{local Markov property} wrt $\mathcal{{G}}$
if each node is conditionally independent of its non-descendants given
its parents. Both properties are equivalent if $\mathbb{{P}}_{X_{1},...,X_{s}}$
has a density%
\footnote{For simplicity we assume that distributions have a density wrt. some
product measure throughout this text.%
} (cf. \cite[Theorem 3.27]{Lauritzen1996}); in this case we say $\mathbb{{P}}_{X_{1},...,X_{s}}$
\emph{is Markov} wrt $\mathcal{{G}}$.
\end{defn}

\begin{defn}[Faithfulness]\label{def:faithful}
$\mathbb{{P}}_{\mathcal{{S}}}$ generated by a SEM $S$ is said to
be \emph{faithful} wrt. $\mathcal{{G_{S}}}$, if 
\[
A\perp_{\text{{d-sep}}}B|C\quad\impliedby\quad A\indep B|C.
\]

\end{defn}
Conveniently the distribution $\mathbb{P}_{\mathcal{S}}$ generated
by a SEM $\mathcal{S}$ is Markov wrt. $\mathcal{G_{\mathcal{S}}}$
(cf. \cite[Theorem 1.4.1]{Pearl2009} for a proof). Hence, if we assume
faithfulness%
\footnote{Intuitively, this is saying that conditional independences are due
to the causal structure and not accidents of parameter values \cite[p. 9]{Spirtes2000}.%
} conditional independences and d-separation properties become equivalent
\begin{align*}
A\perp_{\text{{d-sep}}}B|C\quad & \iff\quad A\indep B|C
\end{align*}

Summing up, we have defined interventional causation in terms of SEMs
and have seen how a SEM gives rise to a DAG. This DAG has two convenient
features. Firstly, the DAG yields a visualisation that allows to easily
grasp missing cause-effect relationships that correspond to missing
directed paths. Secondly, assuming faithfulness d-separation properties
of this DAG are equivalent to conditional independence properties
of the joint distribution. Thus, conditional independences translate
into causal statements, e.\,g. `a variable becomes independent of all
its non-effects given its immediate causes' or `cause and effect are
marginally dependent'. Furthermore, the causal graph $\mathcal{G_{S}}$
can be identified from conditional independences observed in $\mathbb{P}_{\mathcal{S}}$
--- at least up to a so-called \emph{Markov equivalence class,} the
set of graphs that entail the same conditional independences \cite{Verma90}.

\subsection{Optimisation on the Stiefel manifold}\label{sec:optimisation}

The proposed algorithms require optimisation of objective functions over the unit-sphere $O^{d-1}\triangleq\{\bm{x}\in\mathbb{R}^{d}:||\bm{x}||=1\}$.
For generality we view the sphere as a special case of the Stiefel
manifold $V_{d\times p}\triangleq\{\bm{M}\in\mathbb{R}^{d\times p}:\bm{M}^{\top}\bm{M}=\bm{I}_{p\times p}\}$
($p\leq d$) for $p=1$.
Implementing the respective objective functions in Theano \cite{Bergstra2010,Bastien2012}, we use the Python toolbox Pymanopt \cite{townsend2016} to perform optimisation directly on the respective Stiefel manifold using a steepest descent algorithm with standard back-tracking line-search.\footnote{For the experiments presented in this manuscript we set both the minimum step size and gradient norm to $10^{-10}$ (arbitrary choice) and the maximum number of steps to $500$ (generous choice based on preliminary test runs that met the former stopping criteria in much earlier iterations). Our toolbox allows to adjust both parameters.}
This approach is exact and efficient, relying on automated differentiation and respecting the manifold geometry.

\section{The basic MERLiN algorithm}\label{sec:basicmerlin}

We consider a situation in which only a linear combination of observed variables constitutes a meaningful causal variable.
These scenarios naturally arise if only samples of a linear mixture $F_1,...,F_{d'}$ of the underlying causal variables $C_1,...,C_d$ are accessible (cf. Figure~\ref{fig:example-basic}).
Standard causal inference methods cannot infer cause-effect relationships among the causal variables $C_1,...,C_d$ without first undoing the unknown linear mixing (also known as blind source separation).
MERLiN aims to establish a cause-effect relationship among causal variables in a linear network while reconstructing a causal variable at the same time.
In other words, a causal variable is reconstructed by optimising for the statistical properties that imply a certain kind of cause-effect relationship.

In this section we first provide the formal problem description.
We then derive sufficient conditions for the kind of cause-effect relationship MERLiN aims to establish, and discuss assumptions on the linear mixing.
Finally, the basic precision matrix based MERLiN algorithm is introduced, which optimises for these sufficient statistical properties in order to recover a linear causal effect from an observed linear mixture.

\subsection{Formal problem description}\label{sec:problem}

The terminology introduced in Section \ref{sub:CBNs} allows to precisely
state the problem as follows.

\subsubsection{Assumptions}\label{sub:assumptions}

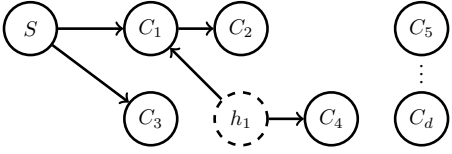
\begin{figure}[tb]
\begin{center}
\-\newline
\begin{tikzpicture}[scale=.8, every node/.style={scale=.8}]
    \node[node] (s) at(-0.5,0) {$S$};
    \node[node] (c1) at(1.5,0) {$C_1$};
    \node[node] (c2) at(3,0) {$C_2$};
    \node[node] (c3) at(1.5,-1.5) {$C_3$};
    \node[nodeh] (h) at(3,-1.5) {$h_1$};
    \node[node] (c4) at(4.5,-1.5) {$C_4$};
    \node[node] (c5) at(6,0) {$C_5$};
    \node at(6,-.65) {$\vdots$};
    \node[node] (cd) at(6,-1.5) {$C_d$};

    \draw[->,d] (s) -- (c1);
    \draw[->,d] (s) -- (c3);
    \draw[->,d] (c1) -- (c2);
    \draw[->,d] (h) -- (c1);
    \draw[->,d] (h) -- (c4);
\end{tikzpicture}\end{center}\caption{Example of a causal graph underlying the described problem scenario (cf. Section~\ref{sec:problem}). $h_{1}$ is a hidden variable.}
\label{fig:example}
\end{figure}

Let $S$ and $C_{1},...,C_{d}$ denote (finitely many) real-valued
random variables. We assume existence of a SEM $\mathcal{{S}}$, potentially
with unobserved variables $h_{1},...,h_{l}$, that induces
$\mathbb{{P}}_{\mathcal{{S}}}=\mathbb{{P}}_{S,C_{1},...,C_{d},h_{1},...,h_{l}}$.
We refer to the corresponding graph $\mathcal{G_{S}}$ as the \emph{true}
\emph{causal graph} and call its nodes \emph{causal variables}. We
further assume that
\begin{itemize}
\item $S$ affects $C_{2}$ indirectly via $C_{1}$,%
\footnote{By saying a variable $X$ causes $Z$ indirectly via $Y$ we imply
(a) existence of a directed path $X\dasharrow Y\dasharrow Z$, and (b) that
there is no directed path $X\dasharrow Z$ without $Y$ on it (this also excludes
the edge $X\to Z$).%
}
\item $\mathbb{{P}}_{\mathcal{{S}}}$ is faithful wrt. $\mathcal{{G_{S}}}$,
\item there are no edges in $\mathcal{{G_{S}}}$ pointing into $S$.
\end{itemize}
In an experimental setting the last condition is ensured by randomising
$S$.%
\footnote{Randomisation corresponds to an intervention: the structural equation
of $S$ is replaced by $S=N_{1}$ where $N_{1}$ is an independent
randomisation variable, e.\,g. assigning placebo or treatment according
to an independent Bernoulli variable.%
} Figure \ref{fig:example} depicts an example of how $\mathcal{{G_{S}}}$
might look.

\subsubsection{Given data}
\begin{itemize}
\item $m$\emph{ iid}%
\footnote{independent and identically distributed%
}\emph{ }samples $\bm{S}=[s_{1},...,s_{m}]^{\top}$ of $S$ and $\bm{F}=[f_{i,j}]_{i=1:m,j=1:d'}$
of $F$ where $F\triangleq[F_{1},...,F_{d'}]^{\top}=\bm{A}C$ is the
observed linear mixture of the causal variables $C\triangleq[C_{1},...,C_{d}]^{\top}$
and $\bm{A}\in\mathbb{{R}}^{d'\times d}$ denotes the unknown mixing matrix
\item the linear combination $\bm{v}\in\mathbb{{R}}^{d'}$ that extracts the causal variable $C_{1}=\bm{v}^{\top}F$ is assumed known
\end{itemize}
That is, we have samples of $S$, $F$, and $C_1$ but not of $C_2,...,C_d$ where $F$ is an unknown linear mixture of $C_1,...,C_d$.

\subsubsection{Desired output}

Find $\bm{w}\in\mathbb{{R}}^{d'}$ such that $aC_{i}=\bm{w}^{\top}F$
where $C_{i}$ is an effect of $C_{1}$ ($i\in\mathbb{{N}}_{2:d},a\in\mathbb{R}\setminus\{0\}$).
In other words, the aim is to recover a causal variable --up to scaling-- that is an effect of $C_1$.
For example, recovery of the causal variable $C_{2}$ is a valid solution.
The factor $a$ reflects the scale indeterminacy that results from the linear mixing, i.\,e., since $\bm{A}$ is unknown the scale of the causal variables cannot be determined unless further assumptions are employed or a priori knowledge is available.

\subsection{MERLiN's strategy}\label{sec:idea}

We are given that there exists at least one causal variable $C_{2}$
that is indirectly affected by $S$ via $C_{1}$. However, we only
have access to samples of the linear mixture $F$ and samples of $S$.
Note the following properties of $C_{2}$:
\begin{itemize}
\item Since $\mathbb{{P}}_{\mathcal{{S}}}$ is faithful wrt. $\mathcal{{G_{S}}}$
it follows that $C_{2}\dep C_{1}$ (and $C_{2}\dep S$).
\item Since $\mathbb{{P}}_{\mathcal{{S}}}$ is Markov wrt. $\mathcal{{G_{S}}}$
it follows that $C_{2}\indep S|C_{1}$.
\end{itemize}
We can derive the following sufficient conditions for
a causal variable to be indirectly affected by $S$ via $C_{1}$.
\begin{claim}
\label{claim:indirecteffect2}Given the assumptions in Section \ref{sub:assumptions}
and a causal variable $Y$. If $Y\dep C_{1}$ and $Y\indep S|C_{1}$,
then $S$ indirectly affects $Y$ via $C_{1}$. In particular, a directed path
from $C_1$ to $Y$, denoted by $C_{1}\dashrightarrow Y$, exists.\end{claim}
\begin{IEEEproof}
From $Y\dep C_{1}$ and $\mathbb{{P}}_{\mathcal{{S}}}$ being Markov
wrt. $\mathcal{{G_{S}}}$ it follows that $Y$ and $C_{1}$ are not
d-separated in $\mathcal{{G_{S}}}$ by the empty set. In $\mathcal{{G_{S}}}$
there must be at least one path $C_{1}\dashrightarrow Y$, $C_{1}\dashleftarrow Y$
or $C_{1}\dashleftarrow X\dashrightarrow Y$ for some node $X$. By
assumption $C_{1}$ is affected by $S$, i.\,e., we have $S\dasharrow C_{1}$
in $\mathcal{G_{S}}$. Hence, in $\mathcal{G_{S}}$ there must be
at least one path $S\dasharrow C_{1}\dasharrow Y$, $S\dasharrow C_{1}\dashleftarrow Y$
or $S\dasharrow C_{1}\dashleftarrow X\dashrightarrow Y$ for some
node $X$. Under the assumption of faithfulness, the latter two cases
contradict $Y\indep S|C_{1}$. Hence, in $\mathcal{G_{S}}$ at least
one path $S\dasharrow C_{1}\dasharrow Y$ exists.

From $Y\indep S|C_{1}$ and $\mathbb{{P}}_{\mathcal{{S}}}$ being
faithful wrt. $\mathcal{{G_{S}}}$ it follows that $Y$ and $S$ are
d-separated in $\mathcal{{G_{S}}}$ by $C_{1}$. That is, given $C_{1}$
every path between $S$ and $Y$ is blocked. In particular, in $\mathcal{{G_{S}}}$
there is no edge $S\to Y$ and no path $S\dasharrow Y$ without $C_{1}$
on it. Hence, $Y$ is indeed indirectly affected by $S$ via $C_{1}$.
\end{IEEEproof}
This leads to our general idea on how to find a linear combination
that recovers a causal effect of $C_{1}$. If MERLiN finds $\bm{w}\in\mathbb{{R}}^{d'}$
such that the following two statistical properties hold true

\begin{flushleft}
\hspace*{2em}(a) $\bm{w}^{\top}F\dep C_{1}$, and\linebreak{}
\hspace*{2em}(b) $\bm{w}^{\top}F\indep S|C_{1}$
\par\end{flushleft}

\noindent then we have identified a candidate causal effect
of $C_{1}$, i.\,e., we have identified a variable such that $S \to C_1 \to \bm{w}^{\top}F$.
Note that conditioning on $S$ cannot unblock a
path that was blocked before since there are no edges pointing into $S$;
conversely the conditional dependence $\bm{w}^{\top}F\dep C_{1}|S$ implies the marginal dependence $\bm{w}^{\top}F\dep C_{1}$.
Hence, MERLiN can also optimise for the following alternative statistical properties

\begin{flushleft}
\hspace*{2em}(a') $\bm{w}^{\top}F\dep C_{1}|S$, and\linebreak{}
\hspace*{2em}(b) $\bm{w}^{\top}F\indep S|C_{1}$
\par\end{flushleft}

\noindent to recover a candidate causal effect of $C_1$.
This reformulation is useful since it allows optimisation of two analogous conditional (in)dependence properties instead of marginal \emph{and} conditional (in)dependence.
Ideally and under mixing assumptions discussed below, optimising
$\bm{w}$ wrt. these statistical properties will indeed recover a
\emph{causal variable}, i.\,e., $\bm{w}^{\top}F=aC_{i}$ ($i\in\mathbb{{N}}_{2:d},a\in\mathbb{R}\setminus\{0\}$),
that is an effect of $C_{1}$. Note that this approach even works
in the presence of hidden confounders.

\subsection{Mixing assumptions}\label{sec:mixing}

MERLiN's strategy presented in the previous section is to optimise a linear combination of the observed linear mixture such that two statistical properties are fulfilled.
However, without imposing further assumptions on the linear mixing it may be impossible to recover the desired causal variable by this procedure.
Here we discuss problems that can occur for arbitrary mixing and specify our mixing assumptions, namely that $\bm{A}$ is an orthogonal $d\times d$ matrix.

In the first place, there may not exist a solution to MERLiN's problem if $\bm{A}$ has
rank less than $d$.%
\footnote{Note that $\bm{A}$ being at least rank $d$ is not a necessary condition,
i.\,e., an effect of $C_{1}$ may be recoverable even in cases where
$\bm{A}$ has rank less than $d$. As an example consider the case
where $C_{2}$ is an effect of $C_{1}$ and $\bm{A}=[\bm{I}_{d\times2},\bm{0}_{d\times(d-2)}]$.
However, the focus is a sufficient condition for the existence of a solution.%
} Hence, assume that $\bm{A}$ has rank $d$ and, for simplicity, that
$\bm{A}$ is a square $d\times d$ matrix. This guarantees existence
of a solution: if the mixing matrix $\bm{A}$ is invertible a solution
to the problem is to recover $C_{2}$ via $\bm{w}=\bm{A}_{2,1:d}^{-1}$.

However, if we only assume $\bm{A}$ to be invertible MERLiN may not
be able to recover (a multiple of) a causal variable $C_{i}$ from
the sought-after statistical properties alone.
The following example demonstrates the problem that arises from the fact that $C_1$ itself is part of the observed linear mixture.

\begin{example}
Assume $S\to C_{1}\to C_{2}\quad C_{3}$ is the true but unknown causal graph,
where the gap indicates that $C_{3}$ is disconnected from all other
variables. Assume all variables are non-degenerate and that the unknown mixing matrix $\bm{A}$ is invertible.
Then, any variable recovered as linear combination from the observed linear mixture $F=\bm{A}C=\bm{A}[C_{1},C_{2},C_{3}]^{\top}$ can be written as\vspace{.5em}

\noindent \vspace{.5em}\resizebox{.99\hsize}{!}{$\left(a\bm{A}_{1,1:d}^{-1}+b\bm{A}_{2,1:d}^{-1}+c\bm{A}_{3,1:d}^{-1}\right)F=aC_{1}+bC_{2}+cC_{3}\triangleq Y_{a,b,c}$}

\noindent for some $a,b,c\in\mathbb{{R}}$.

MERLiN aims to recover the causal variable $C_2$ (up to scaling) by optimising $a,b,c$ such that the statistical properties
\begin{itemize}
    \item $Y_{a,b,c}\dep C_1$ (or equivalently $Y_{a,b,c}\dep C_1|S$), and
    \item $Y_{a,b,c}\indep S|C_{1}$
\end{itemize}
hold true (cf. Section~\ref{sec:idea}).
Indeed $bC_2 = Y_{0,b,0}$ ($b\neq 0$) fulfils these statistical properties and is the desired output.
However, all $Y_{a,0,c}$ ($a,c \neq 0$) likewise fulfil the statistical properties while not being (a multiple of) a causal variable.
\end{example}

This example demonstrates that without imposing further constraints on the linear mixing MERLiN may recover $C_1$ (ensuring the dependence on $S$) with independent variables added on top (ensuring conditional independence of $S$ given $C_1$), e.\,g. $Y_{1,0,1} = C_1 + C_3$ in above example.
Although the sought-after statistical properties hold true for this variable, this is clearly not a desirable output and does not recover a causal variable.

One way to mitigate this situation is to restrict search to the orthogonal
complement $\bm{v}_{\perp}$ of $\bm{v}$. This way, the signal of
$C_{1}$ in the linear mixture $F$ is attenuated. In particular,
if the mixing matrix $\bm{A}$ is orthogonal restricting search to
$\bm{v}_{\perp}$ amounts to complete removal of $C_{1}$'s signal
from $F$.
We therefore assume that $\bm{A}$ is an orthogonal $d\times d$
matrix and restrict the search to $\bm{v}_{\perp}$.
It is then no longer possible to add arbitrary multiples
of $C_{1}$ onto independent variables to introduce the sought-after dependence, i.\,e., the recovery of non-causal variables like $C_1+C_3$ in above example is prevented.

Note that while adding independent variables onto effects is still possible
(e.\,g. consider $Y_{0,1,1}=C_{2}+C_{3}$ in above example), it will
be counter-acted by setting up the objective function accordingly
--- roughly speaking, as we `maximise dependence', then these independent
variables will be suppressed, since they act as noise and reduce dependence.

\subsection{MERLiN$_{\Sigma^{-1}}$: precision matrix magic}\label{sec:alg1}

The basic MERLiN algorithm aims to recover a linear causal effect from an observed linear mixture.
In particular, we assume that the cause-effect relationships $S \to C_1 \to C_2$ between the underlying causal variables $S,C_1$ and $C_2$ are linear with additive Gaussian noise.
In such a linear network, zero entries in the precision matrix imply missing edges in the graph \cite{Lauritzen1996}.
Hence, if $Y$ is a linear effect of $C_{1}$ the precision matrix of the three
variables $S,C_{1}$ and $Y$ is of the form
\[
\Sigma^{-1}\triangleq\Sigma_{S,C_{1},Y}^{-1}=\begin{bmatrix}\star & \star & 0\\
\star & \star & \star\\
0 & \star & \star
\end{bmatrix}
\]
where stars indicate non-zero entries. This implies the partial correlations $\rho_{Y,C_{1}|S}=\star$
and $\rho_{Y,S|C_{1}}=0$ which, in the Gaussian case, amounts to the desired
conditional (in-)dependences (a') $Y\dep C_{1}|S$ and (b) $Y\indep S|C_{1}$
(cf. Section~\ref{sec:idea}) \cite{Baba2004}.

Exploiting this link, the precision matrix based MERLiN$_{\Sigma^{-1}}$
algorithm (cf. Algorithm \ref{alg:alg1}) implements the general idea
presented in Section~\ref{sec:idea} by
maximising the objective function%
\footnote{For numerical reasons one might want to use the approximation $\sqrt{\cdot+\epsilon}\approx|\cdot|$
for small $0<\epsilon\in\mathbb{R}$ to ensure that $f$ is differentiable
everywhere.%
}
\[
f(\bm{w})=\left|\left(\widehat{\Sigma}_{\bm{w}}^{-1}\right)_{2,3}\right|-\left|\left(\widehat{\Sigma}_{\bm{w}}^{-1}\right)_{1,3}\right|
\]
where $\widehat{\Sigma}_{\bm{w}}^{-1}\triangleq\widehat{\Sigma}_{S,C_{1},\bm{w}^{\top}F}^{-1}$  (here we assumed $d\leq m$
and invertibility).
Optimisation is performed over the unit-sphere
$O^{d-2}$ after projecting $\bm{F}$ onto
the orthogonal complement $\bm{v}_{\perp}$.

\begin{algorithm}[tb]
\caption{MERLiN$_{\Sigma^{-1}}$}
\label{alg:alg1}

\textbf{Input:} $\bm{S}\in\mathbb{R}^{m\times1},\bm{F}\in\mathbb{R}^{d\times m},\bm{v}\in\mathbb{R}^{d\times1}$

\textbf{Procedure:}
\begin{itemize}
\item set $\bm{C}:=\bm{F}_{}^{\top}\bm{v}$ and $\bm{F}:=P(\bm{v})\bm{F}\in\mathbb{R}^{(d-1)\times m}$
\item define the objective function for $\bm{w}\in O^{d-2}$ as
\[
f(\bm{w})=\left|\left(\widehat{\Sigma}_{\bm{w}}^{-1}\right)_{2,3}\right|-\left|\left(\widehat{\Sigma}_{\bm{w}}^{-1}\right)_{1,3}\right|
\]
where the empirical precision matrix is
\[
\widehat{\Sigma}_{\bm{w}}^{-1}=\left(\frac{1}{m-1}\left[\bm{S},\bm{C},\bm{F}^{\top}\bm{w}\right]^{\top}\bm{H}_{m}\left[\bm{S},\bm{C},\bm{F}^{\top}\bm{w}\right]\right)^{-1}
\]

\item optimise $f$ as described in Section~\ref{sec:optimisation} to obtain the vector $\bm{w}^* \in O^{d-2}$
\end{itemize}
\textbf{Output: $\bm{w}=P(\bm{v})^{\top}\bm{w}^*\in O^{d-1}$}

\rule[0.5ex]{1\columnwidth}{1pt}

Definitions:
\begin{itemize}
\item $P(\bm{v})$ is the $(d-1)\times d$ orthonormal matrix that accomplishes
projection onto the orthogonal complement $\bm{v}_{\perp}$
\item $\bm{H}_{m}=\bm{I}_{m\times m}-\frac{1}{m}\bm{1}_{m\times m}$ is
the $m\times m$ centering matrix
\end{itemize}
\end{algorithm}

\section{Simulation experiments}\label{sec:simulation}

\subsection{Description of synthetic data}\label{sec:syntheticdata}

$\mathcal{{D}}_{T}^{d\times m}(a,b)$ denotes the synthetic dataset
that is generated by Algorithm \ref{alg:data}. It consists of samples
of an orthogonal linear mixture of underlying causal variables that
follow the causal graph shown in Figure \ref{fig:example}.
The parameters $a$ and $b$ determine the statistical properties of the generated dataset as follows.
The parameter $b$ adjusts the severity of hidden confounding between
$C_{1}$ and $C_{4}$. Note also that the link between $S$ and $C_{2}$
is weaker for higher values of $b$, i.\,e., $\operatorname{corr}(S,C_{2})^{2}=\nicefrac{1}{\left(2+b^{2}+a^{2}\right)}$.
The link
between $C_{1}$ and $C_{2}$ becomes noisier for higher values of
$a$, i.\,e., $\operatorname{corr}(C_{1},C_{2})^{2}=\nicefrac{\left(2+b^{2}\right)}{\left(2+b^{2}+a^{2}\right)}$.
Furthermore, the value of the objective function for recovering $C_2$ is lower for higher values of $a$ since --in the infinite sample limit-- we have
\[\left|\left(\Sigma_{S,C_{1},C_2}^{-1}\right)_{2,3}\right| - \left|\left(\Sigma_{S,C_{1},C_2}^{-1}\right)_{1,3}\right| = \frac{1}{a^2}\]
Hence, these datasets allow to analyse the behaviour of the algorithm for cause-effect relationships of different strengths and its robustness against hidden confounding.

\begin{algorithm}[tb]
\caption{Generating the synthetic dataset $\mathcal{{D}}_{T}^{d\times m}(a,b)$.}
\label{alg:data}

\textbf{Input: $d,m\in\mathbb{N}$, $a,b\in\mathbb{R}$, $T\in\{G,B\}$}

\textbf{Procedure:}
\begin{itemize}
\item generate a random orthogonal$^{a}$ $d\times d$ matrix $\bm{A}$
by Gram-Schmidt orthonormalising a matrix with entries independently
drawn from a standard normal distribution
\item set $\bm{v}^{\top}:=\left(\bm{A}^{-1}\right)_{1,1:d}=\left(\bm{A}^{\top}\right)_{1,1:d}$
\item set $\bm{w}_{G0}^{\top}:=\left(\bm{A}^{-1}\right)_{2,1:d}=\left(\bm{A}^{\top}\right)_{2,1:d}$
\item generate independent mean parameters $\mu_{1},...,\mu_{d},\mu_{h_{1}}$
from $\mathcal{N}(0,1)$
\item generate $m$ independent samples according to the following SEM 
\begin{alignat*}{4}
S & = &  &  &  & N_{0}\\
C_{1} & = &  & \mu_{1}+ &  & N_{1}+S+bh_{1}\\
C_{2} & = &  & \mu_{2}+ & a & N_{2}+C_{1}\\
C_{3} & = &  & \mu_{3}+ &  & N_{3}+S\\
C_{4} & = &  & \mu_{4}+ &  & N_{4}+bh_{1}\\
C_{k} & = &  & \mu_{k}+ &  & N_{k} & (k\in\mathbb{N}_{5:d})
\end{alignat*}
where $(N_{1},...,N_{d})\sim\mathcal{N}(0,1)^{d}$, $h_{1}\sim\mathcal{N}(\mu_{h_{1}},1)$,
and $N_{0}\sim\operatorname{Unif}(\{-1,+1\})$ if $T=B$ or $S\sim\mathcal{N}(0,1)$
if $T=G$
\item arrange the $m$ samples $s_{1},...,s_{m}$ of $S$ in a column vector
$\bm{S}$
\item arrange each sample of $(C_{1},...,C_{d})$ in a column vector and
(pre-)multiply by $\bm{A}$ to obtain the corresponding sample of
$(F_{1},...,F_{d})$
\item arrange the $m$ samples of $(F_{1},...,F_{d})$ as columns of a $d\times m$
matrix $\bm{F}$
\end{itemize}
\textbf{Output: }$\bm{S},\bm{F},\bm{v},\bm{w}_{G0}$

\rule[0.5ex]{1\columnwidth}{1pt}

$^{a}$Since we can ignore scaling, it is not a problem that we in
fact generate an orthonormal matrix.
\end{algorithm}

\subsection{Assessing MERLiN's performance}\label{sub:Performance-measures}

We introduce two performance measures to assess MERLiN's performance
on synthetic data with known ground truth $\bm{w}_{G0}$. Since a
solution can only be identified up to scaling,
we only need to consider the $(d-1)$-sphere $O^{d-1}=\{\bm{x}\in\mathbb{{R}}^{d}:||\bm{x}||=1\}$.
The closer a vector $\bm{w}\in O^{d-1}$ or its negation $-\bm{w}$
is to the ground truth $\bm{w}_{G0}\in O^{d-1}$ the better. This
leads to the performance measure of \textbf{an}gular \textbf{di}stance
\[
\operatorname{andi}_{\bm{w}_{G0}}(\bm{w})\triangleq\min\left(\sphericalangle(\bm{w},\bm{w}_{G0}),\sphericalangle(-\bm{w},\bm{w}_{G0})\right)\in[0,\nicefrac{\pi}{2}]
\]

Another approach is to assess the quality of the recovered $\bm{w}$ by
the probability of obtaining a vector that is closer to $\bm{w}_{G0}$
if chosen uniformly at random on the $(d-1)$-sphere. We define
the \textbf{p}robability \textbf{o}f a \textbf{b}etter \textbf{v}ector
as
\[
\operatorname{pobv}_{\bm{w}_{G0}}(\bm{w})\triangleq\mathbb{{P}}\left[|\bm{w}_{r}\cdot\bm{w}_{G0}|>|\bm{w}\cdot\bm{w}_{G0}|\right]
\]
where $\bm{w}_{r}\sim\operatorname{Unif}(O^{d-1})$ and $d$ is the
dimension of the input vector. This quantity is obtained by dividing
the area of the smallest hyperspherical cap centred at $\bm{w}_{G0}$
that contains $\bm{w}$ or $-\bm{w}$ by half the area of the $(d-1)$-sphere.
The former equals the area of the hyperspherical cap of height \mbox{$h=1-|\bm{w}\cdot\bm{w}_{G0}|$},
the latter equals the area of the hyperspherical cap of height $r=1$.
Exploiting the concise formulas for the area of a hyperspherical cap
with radius $r$ presented in \cite{Li2011} we obtain
\[
\operatorname{pobv}_{\bm{w}_{G0}}(\bm{w})=I_{h(2-h)}\left(\frac{{d-1}}{2},\frac{{1}}{2}\right)
\]
where $h=1-|\bm{w}\cdot\bm{w}_{G0}|$ and $I_{x}(a,b)$ is the regularized
incomplete beta function. It is interesting to note that $I_{x}(\nicefrac{(d-1)}{2},\nicefrac{1}{2})$
is the cumulative distribution function of a $\operatorname{Beta}(\nicefrac{(d-1)}{2},\nicefrac{1}{2})$
distribution such that $\mbox{\ensuremath{|\bm{w}_{r}\cdot\bm{w}_{G0}|^{2}\sim\operatorname{Beta}(\nicefrac{1}{2},\nicefrac{(d-1)}{2})}}$.

For simplicity we drop the ground truth vector $\bm{w}_{G0}$ from
the notation and simply assume that the corresponding ground
truth vector is always the point of reference. Both performance measures are related
inasmuch as $\operatorname{pobv}(\bm{w})=0$ iff $\operatorname{andi}(\bm{w})=0$
and $\operatorname{pobv}(\bm{w})=1$ iff $\operatorname{andi}(\bm{w})=\nicefrac{\pi}{2}$.
However, they capture somewhat complementary information: $\operatorname{andi}(\bm{w})$
assesses how close the vector is in absolute terms, while $\operatorname{pobv}(\bm{w})$
accounts for the increased problem complexity in higher dimensions.

\subsection{Experimental results}\label{sec:simulationresults}

\begin{figure}[tb]
\begin{centering}
\centerline{\includegraphics[width=.97\linewidth]{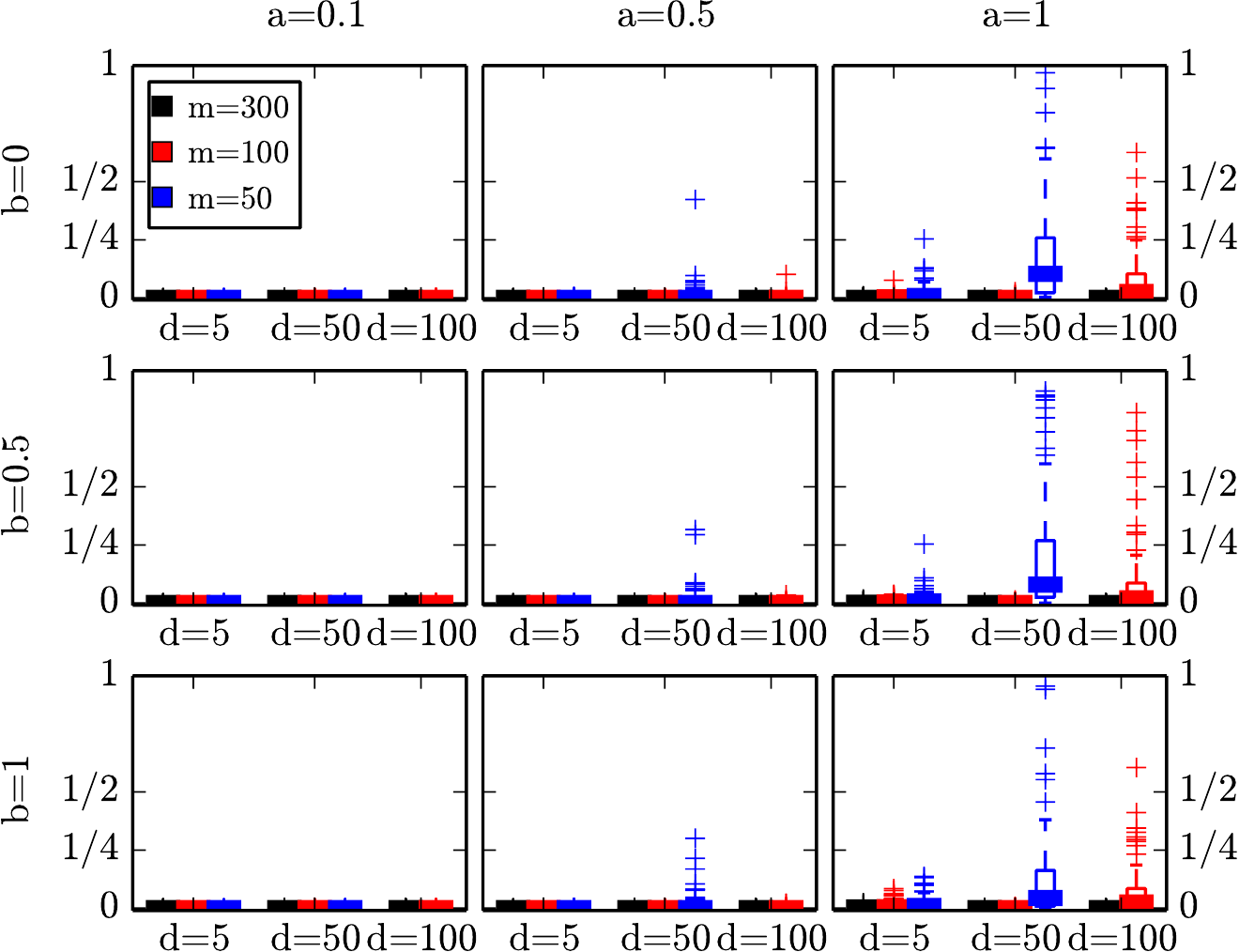}}
\par\end{centering}

\caption{The boxplots summarise the results of $100$ experiments running MERLiN$_{\Sigma^{-1}}$
on datasets $\mathcal{{D}}_{T}^{d\times m}(a,b)$ for $T=G$ (cf. Section
\ref{sec:syntheticdata}). The performance measure $\operatorname{pobv}_{\bm{w}_{G0}}(\bm{w})$
is shown on the $y$-axes and described in Section \ref{sub:Performance-measures}
(low values are good). The box for $d=100,m=50$ is missing since
MERLiN$_{\Sigma^{-1}}$ can only be applied if $d\leq m$.}
\label{fig:gausspobv}
\end{figure}

\begin{figure}[tb]
\begin{centering}
\centerline{\includegraphics[width=.97\linewidth]{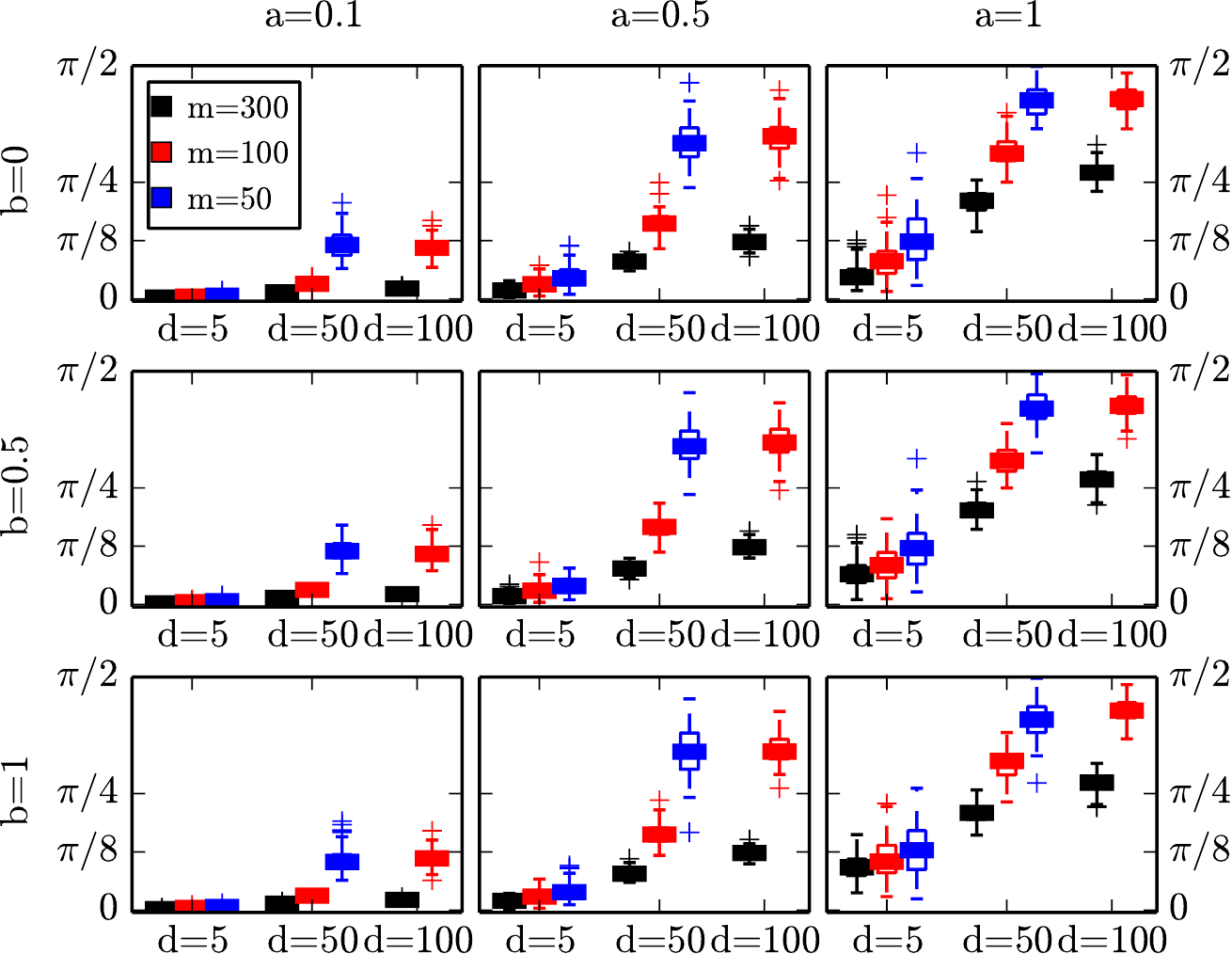}}
\par\end{centering}

\caption{The boxplots summarise the results of $100$ experiments running MERLiN$_{\Sigma^{-1}}$
on datasets $\mathcal{{D}}_{T}^{d\times m}(a,b)$ for $T=G$ (cf.
Section \ref{sec:syntheticdata}). The performance measure $\operatorname{andi}_{\bm{w}_{G0}}(\bm{w})$
is shown on the $y$-axes and described in Section \ref{sub:Performance-measures}
(low values are good). The box for $d=100,m=50$ is missing since
MERLiN$_{\Sigma^{-1}}$ can only be applied if $d\leq m$.}
\label{fig:gaussandi}
\end{figure}

We applied the precision matrix based MERLiN$_{\Sigma^{-1}}$ algorithm
(cf. Algorithm \ref{alg:alg1}) to the synthetic datasets $\mathcal{{D}}_{T}^{d\times m}(a,b)$
described in Section \ref{sec:syntheticdata}. The results of $100$\,runs%
\footnote{For each run we create a new dataset. This is the case for all experiments
on synthetic data. The performance measures $\operatorname{andi}_{\bm{w}_{G0}}(\bm{w})$
and $\operatorname{pobv}_{\bm{w}_{G0}}(\bm{w})$ are always considered
wrt. the corresponding $\bm{w}_{G0}$ of each dataset instance.%
} for different configurations of $d,m,a,b$ are summarised as boxplots
in Figures \ref{fig:gausspobv} and \ref{fig:gaussandi}. Recall that
lower values of $\operatorname{pobv}_{\bm{w}_{G0}}(\bm{w})$
and $\operatorname{andi}_{\bm{w}_{G0}}(\bm{w})$ indicate that
$\bm{w}$ is closer to the ground truth $\bm{w}_{G0}$. We observe
the following:
\begin{itemize}
\item The results for Gaussian ($T=G$) or binary ($T=B$; not shown here)
variable $S$ are indistinguishable.
\item Performance is insensitive to the severity of hidden confounding,
which can be seen by comparing the plots row-wise for the different
values of $b$. This behaviour is expected since $C_{4}\dep S|C_{1}$.
\item Performance decreases with increasing noise level, i.\,e., with increasing
values of $a$. Note that $C_{2}$
is a sum of $C_{1}$ and noise $aN_2$ with variance $2+b^2$ and $a^2$ respectively.
\item The problem becomes harder in higher dimensions, resulting in worse
performance. However, the results for $\operatorname{pobv}_{\bm{w}_{G0}}(\bm{w})$
indicate that even if the solution is not that close to $\bm{w}_{G0}$
in an absolute sense ($\operatorname{andi}_{\bm{w}_{G0}}(\bm{w})$)
the solution is good in a probabilistic sense.
\item More samples increase performance. Especially if the noise level $a$
and the dimension $d$ are not both high at the same time, MERLiN
still achieves good performance on $m=300$ samples (cf. the
results for $a=0.1,\ d=100$ or $a=1,\ d=5$).
\end{itemize}

\section{How to extend MERLiN}\label{sec:extendingmerlin}

In this section we demonstrate how the basic MERLiN algorithm can be extended to enable application to different data modalities by (a) including data specific preprocessing steps into the optimisation procedure (cf. Section~\ref{sec:alg2}), and (b) incorporating a priori domain knowledge (cf. Section~\ref{sec:alg3}).
In particular, we demonstrate this for neuroimaging data, since stimulus-based experiments pose a prototype application scenario for MERLiN for the following reasons.

\begin{enumerate}
    \item In stimulus-based experiments the stimulus $S$ is randomised, meeting the assumption in Section~\ref{sub:assumptions} \cite{weichwald2014}.
    \item Recent work in the neuroimaging community focusses on functional networks, i.\,e., a (linear) combination of activity spread across the brain that is functionally (read \emph{causally}) relevant \cite{vandenHeuvel2010}. Additionally, the recorded activity is often assumed to be a linear combination of underlying cortical variables, as for example in EEG \cite{Nunez2006}.
    \item Simple univariate methods suffice to identify an effect $C_1$ of $S$ \cite[Interpretation rule S1]{Weichwald2015}.
\end{enumerate}

The proposed method can readily be applied and complement the standard analysis procedures employed in such experiments.
More precisely, MERLiN can recover meaningful cortical networks (read \emph{causal variables}) that are causally affected by $C_1$, thereby establishing a cause-effect relationship between two functional brain state features.

\subsection{MERLiN$_{\Sigma^{-1}}^{bp}$: adaptation to EEG data}\label{sec:alg2}

Analyses of EEG data commonly focus on trial-averaged log-bandpower
in a particular frequency band.
Accordingly, we aim at identifying a linear combination of electrode signals such that the trial-averaged log-bandpower of the recovered signal is indirectly affected by the stimulus via another predefined cortical signal.
We demonstrate how to do so by extending the basic MERLiN algorithm to include the log-bandpower computation into the optimisation procedure.

More precisely, we consider EEG trial-data of the form $\widetilde{\bm{F}}\in\mathbb{R}^{d\times m\times n}$
where $d$ denotes the number of electrodes, $m$ the number of trials,
and $n$ the length of the time series $\widetilde{\bm{F}}_{i,j,1:n}$
for each electrode $i\in\mathbb{N}_{1:d}$ and each sample $j\in\mathbb{N}_{1:m}$.%
\footnote{Note that the MERLiN$_{\Sigma^{-1}}$ algorithm takes data of the form $\bm{F} \in \mathbb{R}^{d\times m}$ as input and cannot readily be applied to timeseries data $\widetilde{\bm{F}}\in\mathbb{R}^{d\times m\times n}$.}
The aim is to identify
a linear combination $\bm{w}\in\mathbb{R}^{d\times1}$ such that the
log-bandpower of the resulting one-dimensional trial signals $\bm{w}^{\top}\widetilde{\bm{F}}_{1:d,j,1:n}$
is a causal effect of the log-bandpower of the one-dimensional trial
signals $\bm{v}^{\top}\widetilde{\bm{F}}_{1:d,j,1:n}$. However, since
the two operations of computing the log-bandpower (after windowing)
and taking a linear combination do not commute, we cannot compute
the trial-averaged log-bandpower for each channel first and then apply
the standard precision matrix based MERLiN$_{\Sigma^{-1}}$ algorithm.
Instead, MERLiN$_{\Sigma^{-1}}^{bp}$ has been adapted to the analysis
of EEG data by switching in the log-bandpower computation.

To simplify the optimisation loop we exploit the fact that applying
a Hanning window%
\footnote{We apply a Hanning window in order to keep the feature computation
in line with \cite{grosse2015identification}.%
} and the FFT to each channel's signal commutes with taking a linear
combination of the windowed and Fourier transformed time series. Note
that averaging of the log-moduli ($\log(|\cdot|)$) of the Fourier
coefficients does not commute with taking a linear combination. Hence,
windowing and computing the FFT is done in a separate preprocessing
step (cf. Algorithm \ref{alg:bp}), while the trial-averaged bandpower
is computed within the optimisation loop after taking the linear combination.
Implementation details for the bandpower and precision matrix based
MERLiN$_{\Sigma^{-1}}^{bp}$ algorithm are described in Algorithm
\ref{alg:alg2}. To ease implementation we treat the complex
numbers as two-dimensional vector space over the reals.

\begin{algorithm}[tb]
\caption{Preprocessing for bp algorithm.}
\label{alg:bp}

\textbf{Input: $\bm{S}\in\mathbb{R}^{m\times1},\widetilde{\bm{F}}\in\mathbb{R}^{d\times m\times n},\bm{v}\in\mathbb{R}^{d\times1}$},
the sampling frequency $f_{s}$, and the desired frequency range defined
by $\omega_{1}$ and $\omega_{2}$\textbf{ }

\textbf{Procedure:}
\begin{itemize}
\item set $a:=\lfloor\frac{\omega_{1}n}{f_{s}}\rfloor$, $b:=\lfloor\frac{\omega_{2}n}{f_{s}}\rfloor$,
and $n':=b-a+1$
\item for $i$ from $1$ to $d$, for $j$ from $1$ to $m$

\begin{itemize}
\item center, apply Hanning window and compute FFT, i.\,e., treat $\widetilde{\bm{F}}_{i,j,1:n}$
as a column vector and set $\widetilde{\bm{F}}_{i,j,1:n}:=\bm{TW}\bm{H}_{n}\widetilde{\bm{F}}_{i,j,1:n}$
\end{itemize}
\item extract relevant Fourier coefficients corresponding to $\bm{v}$,
i.\,e., set
\begin{alignat*}{1}
\bm{V}^{\operatorname{Im}}:= & \operatorname{Im}\left(\bm{v}^{\top}\widetilde{\bm{F}}_{1:d,j,a:b}\right)_{j=1:m}\in\mathbb{R}^{m\times n'}\\
\bm{V}^{\operatorname{Re}}:= & \operatorname{Re}\left(\bm{v}^{\top}\widetilde{\bm{F}}_{1:d,j,a:b}\right)_{j=1:m}\in\mathbb{R}^{m\times n'}
\end{alignat*}

\item remove direction $\bm{v}$ from $\widetilde{\bm{F}}$, i.\,e., for $j$
from $1$ to $m$ set
\begin{alignat*}{1}
\bm{F}_{1:(d-1),j,1:n'}^{\operatorname{Im}}:= & \operatorname{Im}\left(P(\bm{v})\widetilde{\bm{F}}_{1:d,j,a:b}\right)\in\mathbb{R}^{(d-1)\times n'}\\
\bm{F}_{1:(d-1),j,1:n'}^{\operatorname{Re}}:= & \operatorname{Re}\left(P(\bm{v})\widetilde{\bm{F}}_{1:d,j,a:b}\right)\in\mathbb{R}^{(d-1)\times n'}
\end{alignat*}
 such that $\bm{F}^{\operatorname{Im}},\bm{F}^{\operatorname{Re}}\in\mathbb{R}^{(d-1)\times m\times n'}$
\end{itemize}
\textbf{Output: $\bm{V}^{\operatorname{Im}},\bm{V}^{\operatorname{Re}}\in\mathbb{R}^{m\times n'}$}
and $\bm{F}^{\operatorname{Im}},\bm{F}^{\operatorname{Re}}\in\mathbb{R}^{(d-1)\times m\times n'}$

\rule[0.5ex]{1\columnwidth}{1pt}

Definitions:
\begin{itemize}
\item $P(\bm{v})$ is the $(d-1)\times d$ orthonormal matrix that accomplishes
projection onto the orthogonal complement $\bm{v}_{\perp}$
\item $\bm{H}_{n}=\bm{I}_{n\times n}-\frac{1}{n}\bm{1}_{n\times n}$ is
the $n\times n$ centering matrix
\item $\bm{W}=\left[\frac{1}{2}\left(1-\cos\frac{2\pi k}{n-1}\right)\right]_{k,l=1:n}$
is the $n\times n$ Hanning window matrix
\item $\bm{T}=\left[\exp\left(-\imath2\pi k\frac{l}{n}\right)\right]_{k,l=1:n}$
is the $n\times n$ FFT matrix
\end{itemize}
\end{algorithm}

\begin{algorithm}[tb]
\caption{MERLiN$_{\Sigma^{-1}}^{bp}$}
\label{alg:alg2}Refer to Algorithm \ref{alg:alg3} and instead use
the objective function
\[
f(\bm{w})=\left|\left(\widehat{\Sigma}_{\bm{w}}^{-1}\right)_{2,3}\right|-\left|\left(\widehat{\Sigma}_{\bm{w}}^{-1}\right)_{1,3}\right|
\]
\end{algorithm}

\subsection{MERLiN$_{\Sigma^{-1}}^{bp+}$: incorporating a priori knowledge}\label{sec:alg3}

Here we demonstrate how to incorporate a priori domain knowledge by modifying the objective function of the MERLiN$_{\Sigma^{-1}}^{bp}$ algorithm.
Utilising a priori knowledge about volume conduction in EEG recordings results in the refined MERLiN$_{\Sigma^{-1}}^{bp+}$ algorithm.

A cortical source projects into more than one EEG electrode. In general,
these volume conduction artefacts might lead to wrong conclusions
about interactions between sources~\cite{Nunez1997}. Imaginary coherency,
as introduced in \cite{Nolte2004}, may help to differentiate volume
conduction artefacts from interactions between cortical sources. To
briefly recap the rationale, we employ the common assumption that the
signals measured at the EEG electrodes have no time-lag to the cortical
signals \cite{Stinstra1998}. The coherency at a certain frequency
of two time series $X$ and $Y$ with Fourier coefficients $x(j),y(j),j\in\mathbb{N}_{1:n}$
is defined as 
\[
\operatorname{coh}_{X,Y}(j)=\frac{\mathbb{E}\left[x(j)y^{*}(j)\right]}{\sqrt{\mathbb{E}\left[x(j)x^{*}(j)\right]\mathbb{E}\left[y(j)y^{*}(j)\right]}}
\]
where $^{*}$ denotes complex conjugation. Next consider the coherency
of $X$ and $Y+X$

\begin{center}
$\operatorname{coh}_{X,Y+X}=\frac{\mathbb{E}\left[x(j)y^{*}(j)\right]+\mathbb{E}\left[x(j)x^{*}(j)\right]}{\sqrt{\mathbb{E}\left[x(j)x^{*}(j)\right]\mathbb{E}\left[\left(y(j)+x(j)\right)\left(y^{*}(j)+x^{*}(j)\right)\right]}}$
\par\end{center}

\noindent and observe that $\mathbb{E}\left[x(j),x^{*}(j)\right]$
is real. This shows that non-zero imaginary coherency $\operatorname{icoh}_{X,Y}(j)\triangleq\operatorname{Im}(\operatorname{coh}_{X,Y}(j))$
cannot be due to volume conduction and indicates time-lagged interaction
between sources since it implies that $\operatorname{Im}(\mathbb{E}\left[x(j)y^{*}(j)\right]) \neq 0$.%
\footnote{Here we exploit the assumption of instantaneous mixing mentioned above.%
}

This a priori knowledge is incorporated in MERLiN$_{\Sigma^{-1}}^{bp+}$
by adapting the objective function to be
\[
f(\bm{w})=\left|\sum_{j=1}^{n'}\operatorname{icoh}(j)\right|\cdot\left|\left(\widehat{\Sigma}_{\bm{w}}^{-1}\right)_{2,3}\right|-\left|\left(\widehat{\Sigma}_{\bm{w}}^{-1}\right)_{1,3}\right|
\]
where $\widehat{\Sigma}_{\bm{w}}^{-1}$ denotes the empirical precision
matrix of the log-bandpower features after taking the linear combination
$\bm{w}$ and $\operatorname{icoh}(j)$ denotes the imaginary coherency
between the signals corresponding to $\bm{v}$ and $\bm{w}$ estimated
as average over all trials (cf. Algorithm \ref{alg:alg3} for details).
While there are several ways of setting up the objective function
we have chosen this multiplicative set-up as it quite naturally captures
the following idea: whenever we find the resulting bandpower to be
dependent on $C_{1}$ we also want to ensure that this is not just
an artefact due to volume conduction. Note that this extension may
also help disentangle true cortical sources, i.\,e., the causal variables,
by avoiding a mixture of distinct sources affected by $C_1$ that have different time-lags
and hence result in lower imaginary coherency.

\begin{algorithm}[tb]
\caption{MERLiN$_{\Sigma^{-1}}^{bp+}$}
\label{alg:alg3}

\textbf{Input: $\bm{S}\in\mathbb{R}^{m\times1},\widetilde{\bm{F}}\in\mathbb{R}^{d\times m\times n},\bm{v}\in\mathbb{R}^{d\times1}$},
the sampling frequency $f_{s}$, and the desired frequency range defined
by $\omega_{1}$ and $\omega_{2}$

\textbf{Procedure:}
\begin{itemize}
\item obtain \textbf{$\bm{V}^{\operatorname{Im}},\bm{V}^{\operatorname{Re}}\in\mathbb{R}^{m\times n'}$}
and $\bm{F}^{\operatorname{Im}},\bm{F}^{\operatorname{Re}}\in\mathbb{R}^{d'\times m\times n'}$
via Algorithm \ref{alg:bp} where $d'=d-1$
\item set \resizebox{.85\hsize}{!}{ $\bm{C}:=\left(\frac{1}{n'}\sum_{j=1}^{n'}\log_{*}\left(\frac{\sqrt{\left(\bm{V}_{i,j}^{\operatorname{Im}}\right)^{2}+\left(\bm{V}_{i,j}^{\operatorname{Re}}\right)^{2}}}{n}\right)\right)_{i=1:m}\in\mathbb{R}^{m\times1}$}\vspace{.5em}\linebreak
(average log-bandpower per trial)
\item define the objective function for $\bm{w}\in O^{d-2}$ as
\[
f(\bm{w})=\left|\sum_{j=1}^{n'}\operatorname{icoh}(j)\right|\cdot\left|\left(\widehat{\Sigma}_{\bm{w}}^{-1}\right)_{2,3}\right|-\left|\left(\widehat{\Sigma}_{\bm{w}}^{-1}\right)_{1,3}\right|
\]
where the empirical precision matrix is
\[
\widehat{\Sigma}_{\bm{w}}^{-1}=\left(\frac{1}{m-1}\left[\bm{S},\bm{C},\bm{D_{w}}\right]^{\top}\bm{H}_{m}\left[\bm{S},\bm{C},\bm{D_{w}}\right]\right)^{-1},
\]
the average log-bandpower per trial depending on $\bm{w}$ is \vspace{.5em}\\\vspace{.5em}\resizebox{.9\hsize}{!}{$\bm{D_{w}}=\left(\frac{1}{n'}\sum_{j=1}^{n'}\log_{*}\left(\frac{\sqrt{\left(\bm{w}^{\top}\bm{F}_{1:d',i,j}^{\operatorname{Im}}\right)^{2}+\left(\bm{w}^{\top}\bm{F}_{1:d',i,j}^{\operatorname{Re}}\right)^{2}}}{n}\right)\right)_{i=1:m}\in\mathbb{R}^{m\times1},$}
and the imaginary coherency $\operatorname{icoh}(j)$ for each frequency
$j\in\mathbb{N}_{1:n'}$ equals \vspace{.5em}\\\vspace{.5em}\resizebox{.9\hsize}{!}{$\frac{\left\langle \bm{V}_{i,j}^{\operatorname{Im}}\cdot\bm{w}^{\top}\bm{F}_{1:d',i,j}^{\operatorname{Re}}-\bm{V}_{i,j}^{\operatorname{Re}}\cdot\bm{w}^{\top}\bm{F}_{1:d',i,j}^{\operatorname{Im}}\right\rangle _{i=1:m}}{\sqrt{\left\langle \left(\bm{V}_{i,j}^{\operatorname{Im}}\right)^{2}+\left(\bm{V}_{i,j}^{\operatorname{Re}}\right)^{2}\right\rangle _{i=1:m}\left\langle \left(\bm{w}^{\top}\bm{F}_{1:d',i,j}^{\operatorname{Im}}\right)^{2}+\left(\bm{w}^{\top}\bm{F}_{1:d',i,j}^{\operatorname{Re}}\right)^{2}\right\rangle _{i=1:m}}}$}
\item optimise $f$ as described in Section~\ref{sec:optimisation} to obtain the vector $\bm{w}^* \in O^{d-2}$
\end{itemize}
\textbf{Output: $\bm{w}=P(\bm{v})^{\top}\bm{w}^*\in O^{d-1}$}

\rule[0.5ex]{1\columnwidth}{1pt}

Definitions:
\begin{itemize}
\item $P(\bm{v})$ is the $(d-1)\times d$ orthonormal matrix that accomplishes
projection onto the orthogonal complement $\bm{v}_{\perp}$
\item $\bm{H}_{m}=\bm{I}_{m\times m}-\frac{1}{m}\bm{1}_{m\times m}$ is
the $m\times m$ centering matrix
\item $\log_{*}$ is the extended log function with $\log_{*}(x)=\log(x),\ x>0$
and $\log_{*}(0)=0$
\item the notation$\langle\cdot\rangle_{i=1:m}$ denotes the empirical mean,
i.\,e., $\langle a_{i}\rangle_{i=1:m}=\frac{1}{m}\sum_{i=1}^{m}a_{i}$\end{itemize}
\end{algorithm}

\section{Experiments on empirical EEG data}

\subsection{Sanity check of the included log-bandpower computation}

We ran simulation experiments with the MERLiN$_{\Sigma^{-1}}^{bp}$ algorithm analogous to those presented in Section~\ref{sec:simulation}.
For this we used datasets $\mathcal{TD}_{T}^{d\times m\times n}(a,b)$ that are generated from $\mathcal{D}_{T}^{d\times m}(a,b)$ with fixed
mixing matrix $\bm{A}=\bm{I}_{d\times d}$ as follows. While $\bm{S},\bm{v},\bm{w}_{G0}$
remain unchanged the $d\times m$ matrix $\bm{F}$ is replaced by
a $d\times m\times n$ tensor $\widetilde{\bm{F}}$ that consists
of $dm$ chunks of randomly chosen real EEG signals of length $n$.
Each signal $\widetilde{\bm{F}}_{i,j,1:n}$ is modified such that
the log-bandpower in the desired frequency band equals $\bm{F}_{i,j}$.

The log-bandpower computation was incorporated into the algorithm in such a way that the optima for MERLiN$_{\Sigma^{-1}}^{bp}$ on $\mathcal{TD}_{T}^{d\times m\times n}(a,b)$ coincide with those for MERLiN$_{\Sigma^{-1}}$ on the corresponding dataset $\mathcal{D}_{T}^{d\times m\times n}(a,b)$; however, the shape of the objective functions is different.
Accordingly and as expected, sanity checks of MERLiN$_{\Sigma^{-1}}^{bp}$ on $\mathcal{TD}_{T}^{d\times m\times n}(a,b)$
show trends for varying parameters $T,d,m,a,b$ similar to those discussed in Section~\ref{sec:simulationresults}.

\subsection{Description of empirical data}

We next evaluate MERLiN with EEG data recorded during a neurofeedback
experiment \cite{GrosseWentrup2014}.%
\footnote{Data was recorded at $121$ active electrodes placed according to
the extended $10$--$20$ system at a sampling frequency of $500$
Hz and converted to common average reference.%
} Subjects in this study were instructed in pseudo-randomised order
to up- or down-regulate the amplitude of $\gamma$-oscillations ($55$--$85$
Hz) in the right superior parietal cortex (SPC). For the feedback
the activity in the SPC was extracted by a linearly-constrained-minimum
variance (LCMV) beamformer \cite{VanVeen1997} that was trained on
$5$ min resting-state EEG data.

Each recording session ($3$ subjects a $2$ sessions referred to
as S1R1, S1R2, S2R1, ...) consists of $60$ trials of $60$ seconds each.
The stimulus variable $S$ is either $+1$ or $-1$ depending on whether
the subject was instructed to up- or down-regulate $\gamma$-power
in the SPC. Electromagnetic artefacts were attenuated as described
in \cite[Section 2.4.1]{GrosseWentrup2014} and the EEG data downsampled
to $250$\,Hz. We are also given the spatial filter $\bm{v}\in\mathbb{R}^{121\times1}$
for each session, i.\,e., the beamformer that was used to extract the
feedback signal. Thus, the data of one session can be arranged as
$\bm{S}\in\{-1,+1\}^{60\times1},\bm{v}\in\mathbb{R}^{121\times1}$
and $\widetilde{\bm{F}}\in\mathbb{R}^{121\times60\times15000}$ where
$\widetilde{\bm{F}}$ contains the timeseries (of length $15000$)
for each channel and trial.

\subsection{Assessing MERLiN's performance}

\begin{figure}[tb]
\centerline{\includegraphics[width=0.95\linewidth]{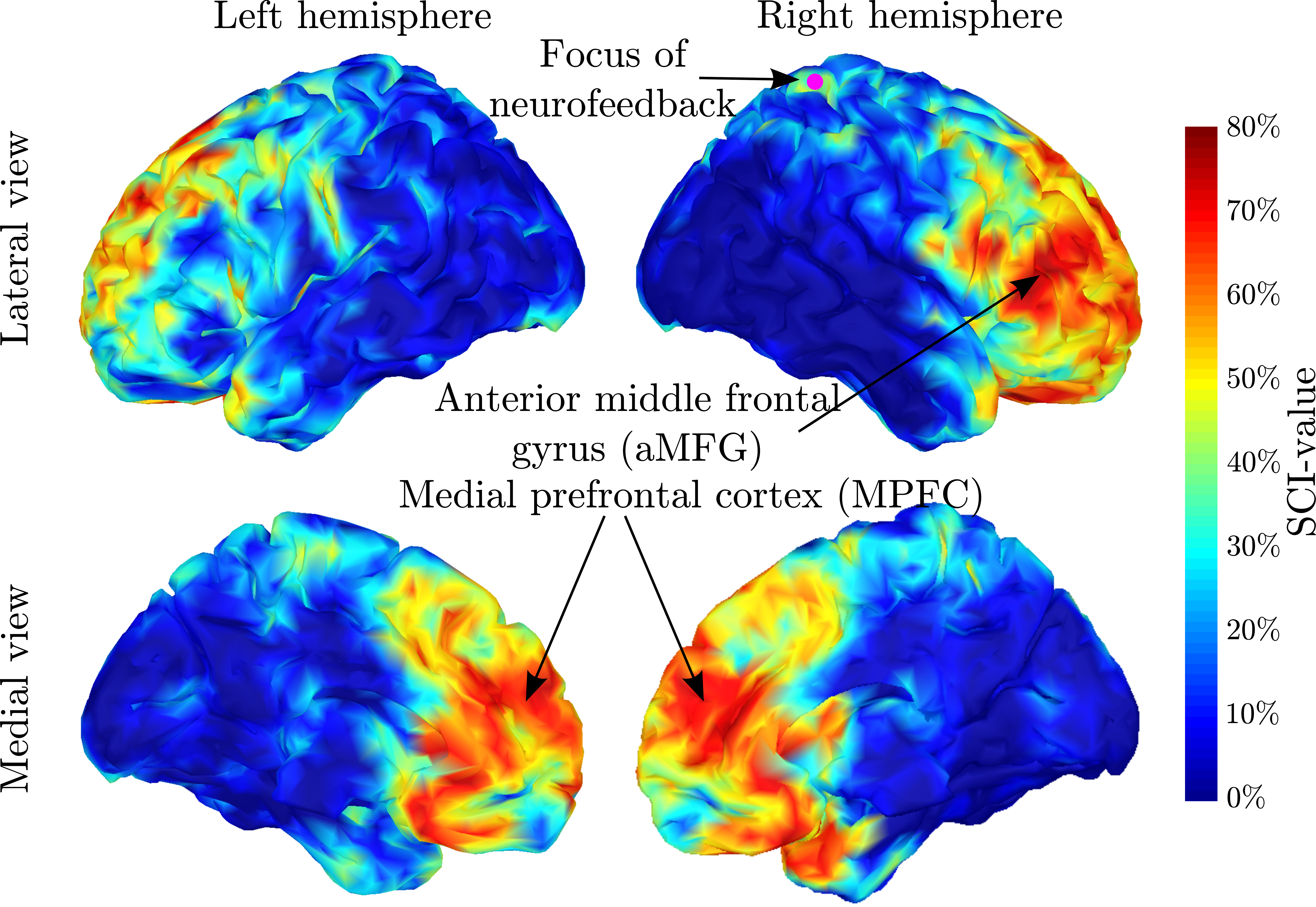}}

\caption{Figure adapted from \cite{grosse2015identification}. The neurofeedback
target area in the right SPC is indicated by a pink circle. The SCI
value denotes the percentage of dipoles within a radius of 7 mm that
were found to be modulated by the SPC. From these results, the authors
inferred the primary targets of the right SPC to be the MPFC and additionally
the right aMFG.}
\label{fig:sciresults}
\end{figure}

MERLiN's performance on these data is assessed by comparing against
results from an earlier exhaustive search approach. The hypothesis
in \cite{grosse2015identification} is that $\gamma$-oscillations
in the SPC modulate $\gamma$-oscillations in the medial prefrontal
cortex (MPFC) and was derived from previous transcranial magnetic stimulation
studies \cite{Chen2013}. In order to test this hypothesis, the signal
of $K\triangleq15028$ dipoles across the cortical surface was extracted
using a LCMV beamformer and a three-shell spherical head model \cite{Mosher1999}.
The SCI algorithm was used to assess for every dipole whether its
$\gamma$-log-bandpower is a causal effect of the $\gamma$-log-bandpower
in the SPC. This analysis confirmed the MPFC as a causal effect of
the SPC (cf. Figure~\ref{fig:sciresults}).

To allow comparison against these results we derive a vector $\bm{g}\in\mathbb{R}^{K\times1}$
that represents the involvement of each cortical dipole in the signal
identified by MERLiN$_{\Sigma^{-1}}^{bp+}$ as the linear combination
$\bm{w}$ of electrode signals. A scalp topography is readily obtained
via $\bm{a}\propto\Sigma\bm{w}$ where the $i^{\text{th}}$ entry
of $\Sigma\bm{w}$ is the covariance between the $i^{\text{th}}$
EEG channel and the source that is recovered by $\bm{w}$ \cite[Equation (7)]{Haufe2014}.
Here $\Sigma$ denotes the subject-specific covariance matrix in the
$\gamma$-frequency band. A dipole involvement vector $\bm{g}$ is
obtained from $\bm{a}$ via dynamic statistical parametric mapping
(dSPM; with the identity as noise covariance matrix) \cite{Dale2000}.
The resulting vectors are expected to be in line with previous findings
and the hypothesis that the MPFC is affected by the SPC.

\subsection{Experimental results}

\begin{figure}[tb]
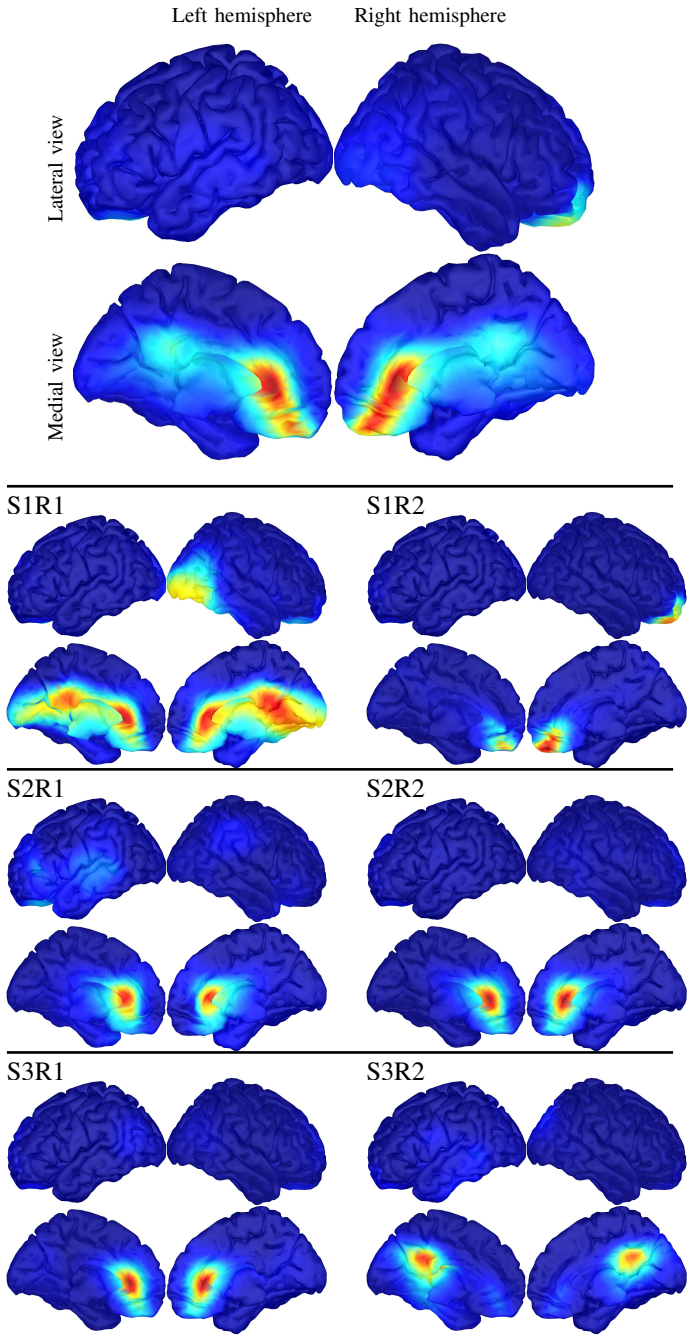

\begin{minipage}[t]{0.49\linewidth}%
Group average

\begin{flushright}
{\footnotesize{}Left hemisphere\quad{}}
\par\end{flushright}{\footnotesize \par}

\begin{flushright}
\rotatebox{90}{\footnotesize\quad Lateral view} \includegraphics[width=0.8\linewidth]{{{F4.group.left.1}}}
\par\end{flushright}%
\end{minipage}%
\begin{minipage}[t]{0.49\linewidth}%
~

\begin{flushleft}
{\footnotesize{}\quad{}Right hemisphere}
\par\end{flushleft}{\footnotesize \par}

\begin{flushleft}
\includegraphics[width=0.8\linewidth]{{{F4.group.right.1}}}
\par\end{flushleft}%
\end{minipage}

\begin{minipage}[t]{0.49\linewidth}%
\begin{flushright}
\rotatebox{90}{\footnotesize\quad Medial view} \includegraphics[width=0.8\linewidth]{{{F4.group.left.2}}}
\par\end{flushright}%
\end{minipage}%
\begin{minipage}[t]{0.49\linewidth}%
\begin{flushleft}
\includegraphics[width=0.8\linewidth]{{{F4.group.right.2}}}
\par\end{flushleft}%
\end{minipage}

\rule[0.5ex]{1\columnwidth}{1pt}

\begin{minipage}[c]{0.24\linewidth}%
S1R1

\includegraphics[width=1\linewidth]{{{F4.1.left.1}}}%
\end{minipage}%
\begin{minipage}[c]{0.24\linewidth}%
\phantom{1}

\includegraphics[width=1\linewidth]{{{F4.1.right.1}}}%
\end{minipage}\hspace{1.5em}%
\begin{minipage}[c]{0.24\linewidth}%
S1R2

\includegraphics[width=1\linewidth]{{{F4.2.left.1}}}%
\end{minipage}%
\begin{minipage}[c]{0.24\linewidth}%
\phantom{1}\includegraphics[width=1\linewidth]{{{F4.2.right.1}}}%
\end{minipage}

\begin{minipage}[c]{0.24\linewidth}%
\includegraphics[width=1\linewidth]{{{F4.1.left.2}}}%
\end{minipage}%
\begin{minipage}[c]{0.24\linewidth}%
\includegraphics[width=1\linewidth]{{{F4.1.right.2}}}%
\end{minipage}\hspace{1.5em}%
\begin{minipage}[c]{0.24\linewidth}%
\includegraphics[width=1\linewidth]{{{F4.2.left.2}}}%
\end{minipage}%
\begin{minipage}[c]{0.24\linewidth}%
\includegraphics[width=1\linewidth]{{{F4.2.right.2}}}%
\end{minipage}

\rule[0.5ex]{1\columnwidth}{1pt}

\begin{minipage}[c]{0.24\linewidth}%
S2R1

\includegraphics[width=1\linewidth]{{{F4.3.left.1}}}%
\end{minipage}%
\begin{minipage}[c]{0.24\linewidth}%
\phantom{1}\includegraphics[width=1\linewidth]{{{F4.3.right.1}}}%
\end{minipage}\hspace{1.5em}%
\begin{minipage}[c]{0.24\linewidth}%
S2R2

\includegraphics[width=1\linewidth]{{{F4.4.left.1}}}%
\end{minipage}%
\begin{minipage}[c]{0.24\linewidth}%
\phantom{1}\includegraphics[width=1\linewidth]{{{F4.4.right.1}}}%
\end{minipage}

\begin{minipage}[c]{0.24\linewidth}%
\includegraphics[width=1\linewidth]{{{F4.3.left.2}}}%
\end{minipage}%
\begin{minipage}[c]{0.24\linewidth}%
\includegraphics[width=1\linewidth]{{{F4.3.right.2}}}%
\end{minipage}\hspace{1.5em}%
\begin{minipage}[c]{0.24\linewidth}%
\includegraphics[width=1\linewidth]{{{F4.4.left.2}}}%
\end{minipage}%
\begin{minipage}[c]{0.24\linewidth}%
\includegraphics[width=1\linewidth]{{{F4.4.right.2}}}%
\end{minipage}

\rule[0.5ex]{1\columnwidth}{1pt}

\begin{minipage}[c]{0.24\linewidth}%
S3R1

\includegraphics[width=1\linewidth]{{{F4.5.left.1}}}%
\end{minipage}%
\begin{minipage}[c]{0.24\linewidth}%
\phantom{1}\includegraphics[width=1\linewidth]{{{F4.5.right.1}}}%
\end{minipage}\hspace{1.5em}%
\begin{minipage}[c]{0.24\linewidth}%
S3R2

\includegraphics[width=1\linewidth]{{{F4.6.left.1}}}%
\end{minipage}%
\begin{minipage}[c]{0.24\linewidth}%
\phantom{1}\includegraphics[width=1\linewidth]{{{F4.6.right.1}}}%
\end{minipage}

\begin{minipage}[c]{0.24\linewidth}%
\includegraphics[width=1\linewidth]{{{F4.5.left.2}}}%
\end{minipage}%
\begin{minipage}[c]{0.24\linewidth}%
\includegraphics[width=1\linewidth]{{{F4.5.right.2}}}%
\end{minipage}\hspace{1.5em}%
\begin{minipage}[c]{0.24\linewidth}%
\includegraphics[width=1\linewidth]{{{F4.6.left.2}}}%
\end{minipage}%
\begin{minipage}[c]{0.24\linewidth}%
\includegraphics[width=1\linewidth]{{{F4.6.right.2}}}%
\end{minipage}

\caption{Spatial pattern of the effect of the SPC as identified by MERLiN$_{\Sigma^{-1}}^{bp+}$.
Group average (first row) and for individual sessions (bottom rows). Each
subplot consists of a lateral (top) and medial (bottom) view of the
left (left) and right (right) hemisphere. (All colorscales from ``blue''
to ``red'' range from $0$ to the largest value to be plotted.)}

\label{fig:brain2}
\end{figure}

We applied MERLiN$_{\Sigma^{-1}}^{bp+}$ several times, i.\,e., with
different random initialisations, to the data of each of the $6$
sessions.%
\footnote{Since there are only $60$ samples per session we decided to select
a subset of $33$ EEG channels distributed across the scalp (again
according to the $10$--$20$ system) after performing the preprocessing
according to Algorithm \ref{alg:bp}. Hence, each run of the
algorithm yields a spatial filter $\bm{w}\in\mathbb{R}^{33\times1}$
and a dipole involvement vector $\bm{g}\in\mathbb{R}^{K\times1}$.%
} We found that the $\gamma$-activation maps $\bm{a}$ obtained for
each spatial filter $\bm{w}$ were (a) rather smooth and similar to
what is typically assumed to be neurophysiologically plausible~\cite{Delorme2012},
and (b) consistent across different initialisations within sessions.
The group average and individual dipole involvement vectors are shown in Figure \ref{fig:brain2}, and Table~\ref{tab:correlations} shows the resulting absolute (partial) correlations between $\gamma$-bandpower in the SPC ($C_1$), the $\gamma$-bandpower of the signal $\bm{w}^\top F$ identified by the MERLiN$_{\Sigma^{-1}}^{bp+}$ algorithm ($C_2$), and the instruction to up- or down-regulate the $\gamma$-bandpower in the SPC ($S$).

Our results are in line with the previous findings (cf. Figure~\ref{fig:sciresults}) inasmuch as they support the hypothesis that the MPFC is a causal effect of the SPC, i.\,e., $S\to C_1 \to C_2$.
We observe the following:
\begin{itemize}
    \item For five out of six sessions and on group average the MPFC shows up as being causally affected by the SPC.
    \item Comparing the marginal correlation $\rho_{S,C_2}$ to the partial correlation $\rho_{S,C_2|C_1}$ suggests that indeed $C_1$ screens off $S$ and $C_2$, which is incompatible with the causal graph $C_1 \gets S \to C_2$. Recall that $C_2 \dep C_1$ and $C_2 \indep S | C_1$ are sufficient to uniquely infer $S\to C_1 \to C_2$ (cf. Section~\ref{sec:idea}).
    \item Unlike the results in \cite{grosse2015identification}, the anterior middle frontal gyrus does not show up in Figure \ref{fig:brain2}.
    \item The parietal/posterior cingulate cortex shows up for sessions S1R1 --here in addition to the MPFC-- and for session S3S2.
\end{itemize}

\begin{figure}[tb]
\begin{center}
\-\newline
\begin{tikzpicture}[scale=.8, every node/.style={scale=.8}]
    \node[node] (s) at(0,0) {$S$};
    \node[node] (c1) at(2,0) {$C_1$};
    \node[node] (c2) at(4,1) {$C_2$};
    \node[node] (c3) at(4,-1) {$C_3$};

    \draw[->,d] (s) -- (c1);
    \draw[->,d] (c1) -- (c2);
    \draw[->,d] (c1) -- (c3);
\end{tikzpicture}
\end{center}\caption{Example causal graph. $C_2$ and $C_3$ are two distinct effects of $C_1$.}
\label{fig:example3}
\end{figure}
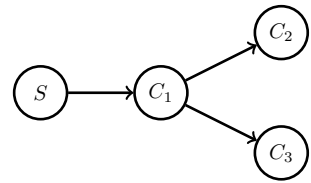

Note that we used MERLiN to recover only one causal variable and hence, that the results are not expected to exactly resemble the exhaustive search results in \cite{grosse2015identification}.
If the true underlying graph is as depicted in Figure~\ref{fig:example3}, then MERLiN may recover any combination $aC_2 + bC_3$ as causal effect of $C_1$.
This may explain both why the anterior middle frontal gyrus does not show up in our analysis --MERLiN recovering only one effect, namely $C_2$ but not $C_3$-- and the lack of intra-subject consistency --slight inter-session differences may lead to recovery of different combinations of effects of $C_1$.
Also note that if the assumption of orthogonal mixing is violated the SPC signal can only be attenuated but not removed (cf. Section~\ref{sec:mixing}).
This may explain the outlier result for session S3R2: The high correlation between $C_1$ and $C_2$ indicates that essentially the SPC signal was recovered, i.\,e., $C_1 \approx C_2$.

\begin{table}[tb]
\caption{Absolute (partial) correlations between $\gamma$-bandpower in the SPC ($C_1$), the $\gamma$-bandpower of the signal $\bm{w}^\top F$ identified by the MERLiN$_{\Sigma^{-1}}^{bp+}$ algorithm ($C_2$), and the instruction to up- or down-regulate the $\gamma$-bandpower in the SPC ($S$).}
\label{tab:correlations}
\begin{center}
\begin{tabular}{c|ccc|c}
Session & $\left|\rho_{S,C_1}\right|$ & $\left|\rho_{C_1,C_2}\right|$ & $\left|\rho_{S,C_2}\right|$ & $\left|\rho_{S,C_2|C_1}\right|$\\
\hline
S1R1 & $0.88$ & $0.36$ & $0.38$ & $0.16$\\
S1R2 & $0.81$ & $0.64$ & $0.51$ & $0.01$\\
S2R1 & $0.34$ & $0.92$ & $0.40$ & $0.23$\\
S2R2 & $0.44$ & $0.55$ & $0.17$ & $0.09$\\
S3R1 & $0.93$ & $0.90$ & $0.83$ & $0.02$\\
S3R2 & $0.88$ & $0.95$ & $0.93$ & $0.67$
\end{tabular}
\end{center}
\end{table}

\section{Discussion}

\subsection{Summary of contributions}

We have proposed a novel idea on how to construct causal variables from observed non-causal variables by explicitly optimising for the statistical properties that imply a certain cause-effect relationship.
This tackles the important problem of causal variable construction, an issue in causal inference that often goes unaddressed and is circumvented by presupposing pre-defined meaningful variables among which cause-effect relationships are to be inferred.
The resulting MERLiN algorithm can recover (or construct) a causal variable from an observed linear mixture that is linearly affected by another given variable.
MERLiN can moreover be extended to enable application to different data modalities by (a) including data specific computation routines into the optimisation procedure, and (b) incorporating further constraints derived from a priori domain knowledge.
We chose to demonstrate this through application to EEG data, since stimulus-based neuroimaging studies are a natural application scenario for MERLiN (cf. Section~\ref{sec:extendingmerlin}).
Results on empirical EEG data indicate that MERLiN can infer meaningful brain state features (read \emph{causal variables}) and establish a cause-effect relationship between two cortical signals.

\subsection{Applications in neuroimaging}

As discussed in Section~\ref{sec:extendingmerlin} interesting application scenarios for MERLiN naturally arise in stimulus-based neuroimaging studies.
MERLiN's fundamental idea is that the construction of causal variables
should explicitly take into account statistical properties that correspond
to causal structure. This supersedes source separation procedures
that often rest on implausible assumptions and are not tailored towards
subsequent causal analyses (e.\,g. ICA in the context of EEG data).
Besides MERLiN's conceptual vantage it is computationally efficient and enables us to bypass both source
localisation (e.\,g. beamforming, dSPM) and an exhaustive search over
$15028$ dipoles.

While we have chosen EEG as an example use case for extended MERLiN algorithms, the extension presented in this manuscript is hoped to serve as a demonstration that will help researchers to adapt the MERLiN algorithm to other neuroimaging modalities.
Future research may focus on extending MERLiN to enable application to functional magnetic resonance imaging data.
This will, due to the high dimensionality compared to the number of samples, again require a modification of the objective function regularising the complexity of the linear combination $\bm{w}$ to avoid perfect recovery of the stimulus variable.

\subsection{Limitations and future research}

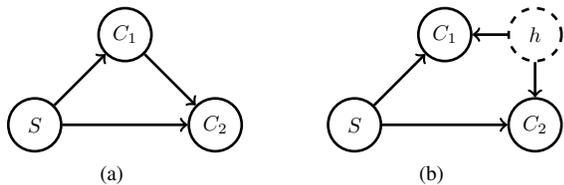
\begin{figure}[tb]
\label{fig:mediation}

\begin{minipage}[t]{0.48\columnwidth}%
\begin{center}
\subfloat[]{
\label{fig:mediation1}

\begin{tikzpicture}[scale=.8, every node/.style={scale=.8}]
    \node[node] (S) at(0,0) {$S$};
    \node[node] (C1) at(1.5,1.5) {$C_1$};
    \node[node] (C2) at(3,0) {$C_2$};
    \draw[->,d] (S) -- (C1);
    \draw[->,d] (C1) -- (C2);
    \draw[->,d] (S) -- (C2);
\end{tikzpicture}\hspace{-1em}
}

\end{center}%
\end{minipage}%
\begin{minipage}[t]{0.48\columnwidth}%
\begin{center}
\subfloat[]{
\label{fig:mediation2}

\begin{tikzpicture}[scale=.8, every node/.style={scale=.8}]
    \node[node] (S) at(0,0) {$S$};
    \node[node] (C1) at(1.5,1.5) {$C_1$};
    \node[node] (C2) at(3,0) {$C_2$};
    \node[nodeh] (h) at (3,1.5) {$h$};
    \draw[->,d] (S) -- (C1);
    \draw[->,d] (h) -- (C1);
    \draw[->,d] (h) -- (C2);
    \draw[->,d] (S) -- (C2);
\end{tikzpicture}\hspace{-1em}
}

\end{center}%
\end{minipage}%

\caption{Example causal graphs where $h$ denotes a hidden confounder.}

\end{figure}

MERLiN tries to identify $\bm{w}^\top F = C_2$ in $S\to C_1 \to C_2$ and rests on the assumption that there is no direct effect $S\to C_2$.
This assumption narrows down the class of causal variables MERLiN can recover, e.\,g. if the true causal graph is as shown in Figure~\ref{fig:mediation1} then MERLiN cannot recover $C_2$.
However, we argue that this is not a strong limitation.
First, in stimulus-based neuroimaging experiments the assumption is likely to be fulfilled if $C_1$ is chosen to be a brain state feature that reflects upstream processing of sensory input, e.\,g. the secondary visual cortex V2 may be assumed to be only indirectly affected by visual stimuli via the primary visual cortex V1.
Second, the MERLiN algorithm is robust in the sense that the statistical properties that it optimises for are sufficient to infer the cause-effect relationships $S \to C_1 \to \bm{w}^\top F$.
In other words, we are on the safe side as long as we refrain from drawing a conclusion if the statistical properties are not met for the identified variable.
Future research may focus on how to recover $C_2$ in scenarios like in Figure~\ref{fig:mediation1}.
This is complicated since the graphs in Figure~\ref{fig:mediation1} and Figure~\ref{fig:mediation2} are Markov equivalent, i.\,e., they entail the same conditional (in-)dependences.
Hence, the cause-effect relationship $C_1 \to C_2$ cannot be reliably inferred from conditional (in-)dependences alone.

MERLiN may be applied in the $d > m$ (high dimension and small sample) setting if an additional regularisation term penalizes the complexity of the linear combination $\bm{w}$.
This leads to the following more general form of the objective function
\[
f(\bm{w})=(1-\lambda)\left|\left(\widehat{\Sigma}_{\bm{w}}^{-1}\right)_{2,3}\right|-\lambda \left|\left(\widehat{\Sigma}_{\bm{w}}^{-1}\right)_{1,3}\right| - \operatorname{complexity}(\bm{w})
\]
where the additional parameter $\lambda \in [0,1]$ may enable to improve performance by weighing dependence/conditional independence depending on the problem structure at hand.

Another limitation is that the MERLiN algorithm presented in this manuscript only works for linear networks, i.\,e., it fails for non-linear cause-effect relationships.
This may not be a strong limitation for neuroimaging applications given that there is empirical evidence that the relations found in EEG and functional magnetic resonance imaging are predominantly linear \cite{muller2003linear,naselaris2011encoding}.
Nevertheless, future research will focus on extending MERLiN to non-linear cause-effect relationships, with preliminary results already available \cite{weichwald2016recovery}.

Future research may also investigate possibilities to assess the statistical significance and uncertainty associated with the linear combination identified by MERLiN.
The former may be accomplished by a permutation scheme that involves running the optimisation for each permutation.
However, it cannot be accomplished by standard significance tests for (conditional) dependence, since an optimisation procedure is employed in obtaining the variables being tested, and this procedure must be corrected for when determining the threshold for significance.
The latter may be accomplished by bootstrap techniques.

\subsection{Disentangling multiple effects}

MERLiN cannot unambiguously recover multiple effects separately (e.\,g. $A,B$ or $C$ in Figure~\ref{fig:example2}) as opposed to any linear combination of those effects that all satisfy the sought-after statistical properties (e.\,g. $aA+bB+cC$ in Figure~\ref{fig:example2}).
However, incorporating a priori knowledge
as demonstrated in Section \ref{sec:alg3} can mitigate this. When analysing EEG data, for
instance, one could a priori exclude spatial filters that are neurophysiologically
implausible and run optimisation over the complement set instead of
the whole unit-sphere.

\subsection{Faithfulness}

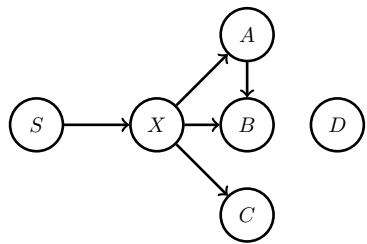
\begin{figure}[tb]
\begin{center}
\-\newline
\begin{tikzpicture}[scale=.8, every node/.style={scale=.8}]
    \node[node] (s) at(-0.5,0) {$S$};
    \node[node] (c1) at(1.5,0) {$X$};
    \node[node] (c2) at(3,1.5) {$A$};
    \node[node] (c3) at(3,0) {$B$};
    \node[node] (c4) at(3,-1.5) {$C$};
    \node[node] (c5) at(4.5,0) {$D$};

    \draw[->,d] (s) -- (c1);
    \draw[->,d] (c1) -- (c2);
    \draw[->,d] (c1) -- (c3);
    \draw[->,d] (c1) -- (c4);
    \draw[->,d] (c2) -- (c3);
\end{tikzpicture}\end{center}\caption{Example causal graph for which it is supposed that the indirect
($X\to A\to B$) and direct ($X\to B$) effects of $X$ on $B$ cancel.}
\label{fig:example2}
\end{figure}

While the faithfulness assumption remains untestable we are unlikely to encounter violations in practice, e.\,g. we
can show that faithfulness holds almost surely if the causal relationships
are linear \cite{Meek1995}. Multivariate causal inference methods
may be robust against certain violations of faithfulness, and hence
offer an alternative to such arguments. MERLiN, for example, is able
to identify cause-effect relationships in unfaithful scenarios that
cannot be revealed by classical univariate approaches. Consider the
graph shown in Figure \ref{fig:example2} and suppose that the indirect
($X\to A\to B$) and direct ($X\to B$) effects of $X$ on $B$ cancel,
i.\,e., $X\indep B$ wrt. the resulting and unfaithful joint distribution.
In this example, univariate methods cannot infer the existence of
the edge $X\to B$, while MERLiN can in principle determine that $B$
is part of the revealed linear combination and as such directly affected
by $X$.
The link to faithfulness prompts further research on multivariate
methods and variants of the faithfulness assumption. Furthermore,
it stresses the importance of causal variable construction.

\balance\bibliographystyle{ieeetr}
\bibliography{references}

\end{document}